\documentclass[prd,preprintnumbers,twocolumn,eqsecnum,floatfix,a4paper,nofootinbib]{revtex4}
\usepackage{color}
\usepackage{calc}
\usepackage{amsmath,amssymb,graphicx}
\usepackage{amssymb,amsmath}
\usepackage{tensor}
\usepackage{bm}
\usepackage{microtype}
\usepackage{booktabs}
\usepackage{times}
\usepackage[varg]{txfonts}
\usepackage[colorlinks, pdfborder={0 0 0}]{hyperref}
\usepackage{subfigure}
\definecolor{LinkColor}{rgb}{0.75, 0, 0}
\definecolor{CiteColor}{rgb}{0, 0.5, 0.5}
\definecolor{UrlColor}{rgb}{0, 0, 0.75}
\hypersetup{linkcolor=LinkColor}
\hypersetup{citecolor=CiteColor}
\hypersetup{urlcolor=UrlColor}
\allowdisplaybreaks
\DeclareFontFamily{OT1}{pzc}{}
\DeclareFontShape{OT1}{pzc}{m}{it}{<-> s * [1.10] pzcmi7t}{}
\DeclareMathAlphabet{\mathpzc}{OT1}{pzc}{m}{it}

\newcommand{\h}{\mathpzc{h}}

\newcommand{\hlm}{\mathpzc{h}_{\ell m}}

\newcommand{\Ylm}{{Y}^{-2}_{\ell m}}

\newcommand{\hc}{h_\times}
\newcommand{\hp}{h_+}
\newcommand{\Fc}{F_\times}
\newcommand{\Fp}{F_+}

\newcommand{\blambda}{\bm{\lambda}}

\newcommand{\Mo}{M_{\odot}}
\newcommand{\FFe}{\mathrm{FF}_\mathrm{eff}}
\newcommand{\FF}{\mathrm{FF}}

\newcommand{\rhoopt}{\rho_\mathrm{opt}}
\newcommand{\rhosubopt}{\rho_\mathrm{subopt}}

\newcommand*{\skymapscale}{0.5}
\newcommand*{\paramestscale}{0.455}

\begin{document}

\newcommand{\be}{\begin{equation}}
\newcommand{\ee}{\end{equation}}
\newcommand{\ber}{\begin{eqnarray}}
\newcommand{\eer}{\end{eqnarray}}
\def\bea{\begin{eqnarray}}
\def\eea{\end{eqnarray}}
\newcommand{\etal}{\emph{et al}}

\title{Gravitational-wave observations of binary black holes: Effect of non-quadrupole modes}
\author{Vijay Varma}
\author{Parameswaran~Ajith}
\affiliation{International Centre for Theoretical Sciences, Tata Institute of Fundamental Research, Bangalore 560012, India}
\author{Sascha Husa}
\author{Juan Calderon Bustillo}
\affiliation{Departament de F\'isica, Universitat de les Illes Balears and Institut d'Estudis Espacials de Catalunya, Crta. Valldemossa km 7.5, E-07122 Palma, Spain}
\author{Mark Hannam}
\author{Michael P\"urrer}
\affiliation{School of Physics and Astronomy, Cardiff University, Queens Building, CF24 3AA, Cardiff, United Kingdom}

\begin{abstract}
We study the effect of non-quadrupolar modes in the detection and parameter estimation of gravitational waves (GWs) from non-spinning black-hole binaries. We evaluate the loss of signal-to-noise ratio and the systematic errors in the estimated parameters when one uses a quadrupole-mode template family to detect GW signals with all the relevant modes, for target signals with total masses $20 M_\odot \leq M \leq 250 M_\odot$ and mass ratios $1 \leq q \leq 18$. Target signals are constructed by matching numerical-relativity simulations describing the late inspiral, merger and ringdown of the binary with post-Newtonian/effective-one-body waveforms describing the early inspiral. We find that waveform templates modeling only the quadrupolar modes of the GW signal are sufficient (loss of detection rate $< 10\%$) for the detection of GWs with mass ratios $q\leq4$ using advanced GW observatories. Neglecting the effect of non-quadrupole modes will introduce systematic errors in the estimated parameters. The systematic errors are larger than the expected $1\,\sigma$ statistical errors for binaries with large, unequal masses ($q\gtrsim4, M \gtrsim 150 M_\odot$), for sky-averaged signal-to-noise ratios larger than $8$. We provide a summary of the regions in the parameter space where neglecting non-quadrupole modes will cause unacceptable loss of detection rates and unacceptably large systematic biases in the estimated parameters. 
\end{abstract}
\preprint{LIGO-P1400095-v3}
\maketitle
\section{Introduction and Summary}
The first direct detection of gravitational waves (GWs) from a ground-based observatory is expected to happen in the next few years. A worldwide network of laser interferometric GW detectors comprising of Advanced LIGO in the USA, Advanced Virgo in Italy, the upcoming KAGRA in Japan and possibly a third LIGO detector in India will soon be operating in conjunction. One of the most promising sources of GWs for ground based detectors is the coalescence (inspiral, merger and ringdown) of binary black holes (BBHs). These systems lose energy and angular momentum through gravitational radiation and inspiral toward each other until they eventually coalesce.

The search for GW signals from BBHs is performed by \emph{matched filtering}, which uses template models of the expected signal to comb through the data from the detector. However, the GW signal is buried deeply in noise and the ability of matched filtering to detect the signal and to determine the properties of the source depends crucially on how accurately the template models the signal present in the data. If the template is a poor approximation of the true signal, this can affect matched filtering in two ways: (i) it can reduce the signal-to-noise ratio (SNR), potentially causing non-detection, (ii) even if the signal is detected, the estimated parameters of the source can be systematically biased. As the goal of GW astronomy is not just detection of GWs but to extract astrophysical information about the source, the waveform templates should be not only \emph{effectual} in detection  (small loss in the SNR), but also \emph{faithful} in parameter estimation (small systematic biases)~\cite{DIS98}.

Gravitational waves, being a tensor field, can be decomposed in terms of the spin $-2$ weighted spherical harmonic basis functions $\Ylm$. GW searches in the past~\cite{Abadie:2011kd,Aasi:2012rja} employed templates~\cite{Buonanno:2007pf,Pan:2011gk,Ajith:2007kx,Ajith:2007xh,Ajith:2009bn} that consisted of only the dominant modes ($\ell=2,m=\pm2$) in this expansion. While quadrupole modes are indeed the dominant modes, actual signals will in general have contributions from all the modes and the sub-dominant modes may play an important role in detection and parameter estimation of BBHs, particularly for binaries with high mass ratios and those highly inclined with respect to the detector.
\subsection{Summary of past studies}
The effect of non-quadrupole modes in the context of post-Newtonian (PN) inspiral waveforms (which appears as higher order corrections to the amplitude) was first studied by Sintes \& Vecchio~\cite{Sintes:1999cg} and explored in detail by Van Den Broeck \& Sengupta~\cite{VanDenBroeck:2006ar,VanDenBroeck:2006qu}. They found that the higher order corrections typically decrease the amplitude of the PN waveforms, causing a reduction in the SNR. Nevertheless, the high frequency content introduced by the higher harmonics (the $m > 2$ modes) can significantly reduce the statistical errors in the parameter estimation for binaries with large ($M \gtrsim 50 M_\odot$) masses, observed by advanced ground-based detectors~\cite{VanDenBroeck:2006ar}. However, in this mass range, the effect of merger-ringdown becomes non-negligible; in the mass range ($M \lesssim 15 M_\odot$) where it suffices to consider only the inspiral stage, recent studies have shown that the effect of higher harmonics is marginal~\cite{Cho:2012ed,O'Shaughnessy:2013vma,O'Shaughnessy:2014dka}. On the other hand, in the context of the space-borne detector LISA, higher harmonics are expected to bring significant reduction in statistical errors~\cite{Arun:2008xf,Arun:2007hu}. 

While the earlier work discussed above considered only the inspiral part of the GW signal, in the recent past, when numerical relativity (NR) simulations have become routine, several groups have investigated the effect of sub-dominant modes in the detection of BBHs using waveforms describing the complete inspiral, merger and ringdown stages of the coalescence. Pekowsky \etal~\cite{Pekowsky:2013hm} studied how well quadrupole-mode waveforms match waveforms that include sub-dominant modes for different orientations of the binary with respect to the detector. The matches were evaluated by using NR waveforms as both target and template waveforms at the same point in the parameter space. For non-spinning BBHs with mass ratios $q \equiv m_1/m_2 \leq 15$ and total masses $M \equiv m_1+m_2 > 100\Mo$ they find that the match (that was not maximized over the masses of the templates) can be lower than $0.97$ for up to $65\%$ of source orientations. However, orientations that correspond to the least matches also correspond to those with least intrinsic luminosity, therefore the effect of sub-dominant modes is suppressed. While Pekowsky \etal\ calculated matches using the same parameters for the target and template waveforms, actual GW searches employ a template bank over which the match is maximized. Brown \etal~\cite{Brown:2013hm} studied the same problem using a template bank of quadrupole-mode-only effective-one-body waveforms calibrated to numerical relativity simulations (EOBNRv2)~\cite{Pan:2011gk}. This study, which employed EOBNRv2 waveforms that include sub-dominant modes as the ``target signals'', concluded that for non-spinning BBHs with component masses $3\Mo \leq m_1,m_2 \leq 25 \Mo$, the maximum loss in the detection rate for a binary with given mass parameters (after averaging over other parameters) is less than $\sim 10\%$. While Brown \etal's investigation considered only binaries with $m_1,m_2 \leq 25 \Mo$, non-quadrupole modes are expected to be more important for binaries with even higher masses. Capano \etal~\cite{Capano:2013hm} recently extended this study to $m_1,m_2 \leq 200 \Mo$. While the study by Brown \etal\ characterized only the loss of SNR of the quadrupole-mode template bank, Capano \etal\ studied, in addition to this, the effect of non-quadrupole modes on the ``$\chi^2$'' signal-based veto. They also compared the efficiency of a search employing ``full-mode'' templates with a search using only quadrupole-mode templates after considering the increased false alarm probability (due to the increase in the number of templates). They conclude that, a search employing a full-mode template bank will actually result in a worse sensitivity than one employing a quadrupole-mode-only bank for $q\lesssim 4$ due to the increase in threshold SNR required to keep the false alarm probability fixed. For binaries with $q > 4$, inclusion of higher modes in the waveform templates can produce a moderate improvement in the detection volume. 

While the studies mentioned above investigated the effect of non-quadrupole modes on the \emph{detection} of GWs, Littenberg \etal~\cite{Littenberg:2012uj} studied the systematic errors in the estimated parameters and compared them against the expected statistical errors using a parameter estimation algorithm employing Markov-Chain Monte-Carlo (MCMC) technique. Because of the computational cost of the MCMC algorithm, the study had to be restricted to a few sample points in the parameter space. They concluded that, for binaries in the range $1 \leq q \leq 6$ and $M < 60 M_\odot$ with a fixed inclination angle $\iota = \pi/3$, the systematic errors introduced by neglecting non-quadrupole modes are smaller than the expected statistical errors at SNR $\lesssim 12$. However, for larger masses ($M = 120 M_\odot, q = 6, \iota = \pi/3$), they have found that neglecting higher modes will cause systematic biases larger than the statistical errors at SNR $\simeq 12$. 

\subsection{Summary of this study}

\begin{figure}[bth]
\begin{center}
\includegraphics[width=3.3in]{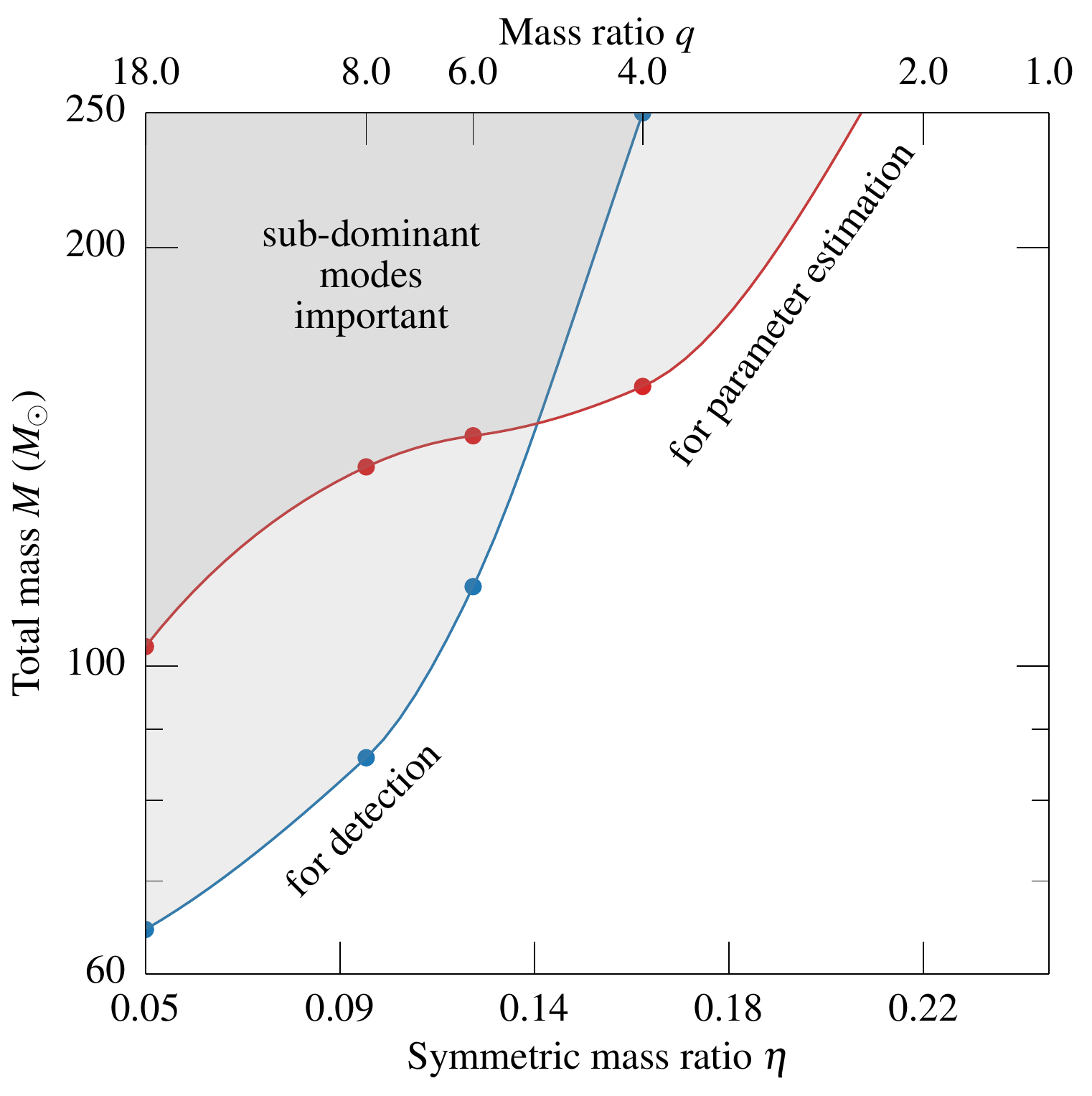}
\caption{This plot summarizes the region in the parameter space of non-spinning black-hole binaries where contributions from non-quadrupole modes are important for GW detection and parameter estimation. The bottom horizontal axis reports the symmetric mass ratio of the binary while the top horizontal axis shows the mass ratio. The vertical axis reports the total mass. Shaded areas show the regions in the parameter space where the loss of detection rate due to neglecting non-quadrupole modes is larger than $10\%$ and/or the systematic bias in the estimated parameters is larger than the expected statistical errors for a sky-averaged SNR of 8.}
\label{fig:summary_fig}
\end{center}
\end{figure}
%
While the study by Pekowsky \etal\ uses NR waveforms as target signals, it was rather incomplete in taking into account all the relevant aspects of the GW searches. The studies by Brown \etal\ and Capano \etal, while being exhaustive in considering the relevant aspects of the GW searches, use a semi-analytical waveform family (EOBNRv2, which models only 4 sub-dominant modes) to describe the target signals. Here we supplement the earlier work by revisiting this problem: As our target signals, we use ``hybrid waveforms'' containing all the relevant modes (with $\ell <= 4$). The hybrid waveforms are constructed by matching NR simulations describing the late inspiral, merger and ringdown of the binary with PN/EOB waveforms describing the early inspiral. We consider the effective volume of a search (1 $-$loss of detection rate) using quadrupole-mode template banks after averaging over all the relative inclinations of the binary with respect to the detector. Our results are broadly in agreement with those obtained by Capano \etal. In addition to the detection aspect, we also study the effect of sub-dominant modes in parameter estimation by characterizing the systematic errors in estimating the binary parameters using a quadrupole-only template family. While Littenberg \etal\ studied the systematic and statistical errors at a handful of points in the parameter space (assuming fixed orientation for target binaries), we compare the systematic biases averaged over all angles describing the relative orientation of the binary and compare them against the sky-averaged statistical errors. While Littenberg \etal\ used an MCMC algorithm to compute statistical and systematic errors, we compute the systematic errors by maximizing the match of the quadrupole-only template bank with the target signals including all modes. Statistical errors are computed using the Fisher matrix formalism employing quadrupole-only templates. Wherever comparisons are possible, our results are broadly in agreement with those of Littenberg \etal. 
 
We consider non-spinning BBHs with total masses $20\Mo \leq M \leq 250\Mo$ and mass ratios $1 \leq q \leq 18$. Hybrid waveforms with $q \leq 8$ are constructed by matching NR waveforms computed by the SpEC code~\cite{Ossokine:2013zga,Hemberger:2012jz,Szilagyi:2009qz,Boyle:2009vi,Scheel:2008rj,Boyle:2007ft,Scheel:2006gg,Lindblom:2005qh,Pfeiffer:2002wt,SpEC,Mroue:2012kv,Mroue:2013xna, Buchman:2012dw}, kindly made public by the SXS collaboration~\cite{SXS-Catalog}, with PN/EOB waveforms describing the early inspiral. The phase of the inspiral waveforms is computed in the EOB method and the amplitude of the spherical harmonics modes are computed in the PN approximation accurate to 3PN order.  For $q=18$, the NR simulation is performed using the BAM code~\cite{Bruegmann:2006at,Husa:2007hp}. We include all modes up to $\ell = 4~(m = -\ell$ to $\ell$, except $m = 0$) in the hybrid waveforms. As template waveforms (quadrupole mode only) we use EOBNRv2~\cite{Pan:2011gk}, an effective-one-body waveform calibrated to numerical relativity simulations. The match between the hybrid waveforms and quadrupole mode templates is maximized over the two mass parameters of the templates by the Nelder-Mead down-hill 
simplex algorithm.

Figure~\ref{fig:summary_fig} provides an executive summary of the main results. The plot shows the region in the parameter space where contribution from non-quadrupole modes are important for detection and parameter estimation. The horizontal axis reports the symmetric mass ratio $\eta$ of the binary and the vertical axis reports the total mass $M$. Shaded areas show the regions in the parameter space where the loss of detection rate due to neglecting non-quadrupole modes is larger than $10\%$ and/or the systematic bias in the estimated parameters (averaged over all orientations of the binary) are larger than the expected statistical errors for a SNR of 8 (averaged over all sky-locations and orientations of the binary). We have found that neglecting non-quadrupole modes causes large systematic errors (larger than the corresponding statistical errors) in the estimation of $M$, while the estimation of $\eta$ is largely unaffected by this. 

The rest of this paper is organized as follows: Sec.~\ref{sec:gw_intro} gives a brief introduction to the observation of GWs from BBHs and introduces the figures of merit used for this study. Section~\ref{sec:methods} provides further details of the methodology, such as details of the NR simulations, construction of the hybrid waveforms, the choice of the template family and the detector model used in this study. Section~\ref{sec:results} discusses our results. This is followed by some concluding remarks which also lists the limitations of this study and possible future work. Throughout this paper, we follow the convention $G=c=1$. We refer to waveforms that include contributions from sub-dominant modes ($\ell = 2$ to 4, $m = -\ell ~ \mathrm{to} ~ \ell$, except the $m = 0$) as ``full'' waveforms, and waveforms that include only quadrupole modes ($\ell = 2, m = \pm 2$) as ``quadrupole'' waveforms. 

\section{Observing gravitational waves from binary black holes}
\label{sec:gw_intro}

\begin{figure*}[bth]
\begin{center}
\includegraphics[width=5.8in]{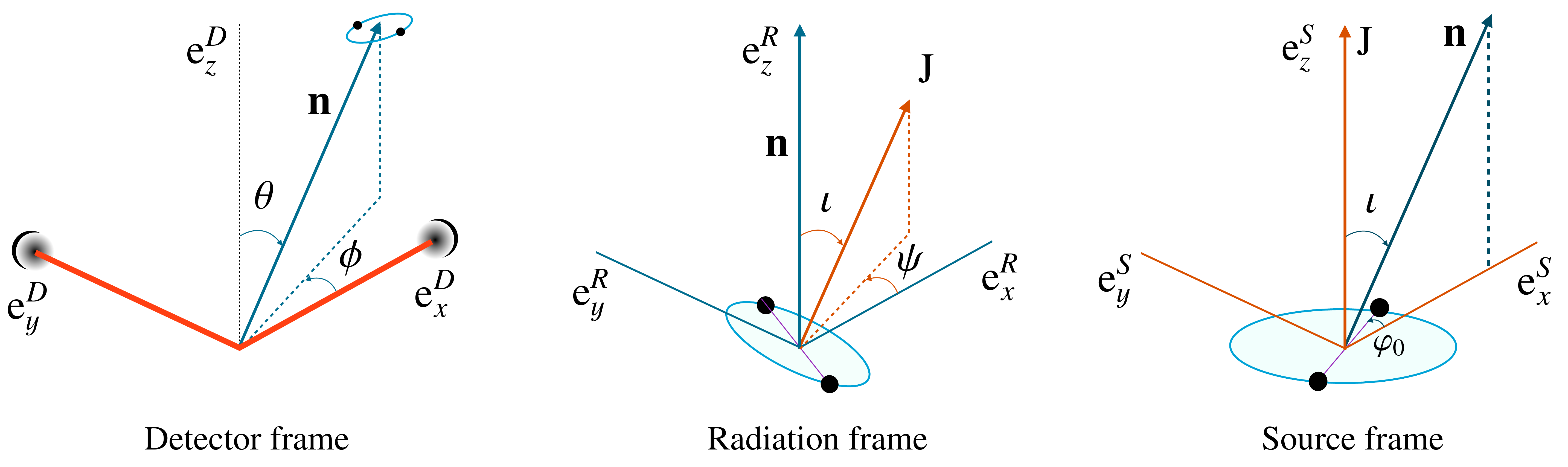}
\caption{\emph{Detector frame:} The two orthogonal arms of the interferometer form the $x$ and $y$ axes in the detector frame while the $z$ axis is defined by the right circular convention.  Angles $\theta$ and $\phi$ denote the polar and azimuth angles of the binary in the sky measured in the detector frame. These angles fix the location of the source in the sky, with respect to the detector. \emph{Radiation frame:} The $z$ axis of the radiation frame is defined by the line-of-sight vector $\mathbf{n}$ from the detector to the source so that the $x-y$ plane is the plane perpendicular to $\mathbf{n}$ (the ``sky''); $x$ axis is defined by the $x$ axis of the detector projected onto the sky. Angles $\iota$ and $\psi$ denote the polar and azimuth angles of the total angular momentum vector $\mathbf{J}$ of the binary in the radiation frame. These angles fix the relative orientation of the binary with respect to the detector. \emph{Source frame:} The $z$ axis of the source frame is defined by the total angular momentum vector $\mathbf{J}$ of the binary and the $x$ axis is defined by the projection of the line of sight onto the binary plane. The angle $\varphi_0$ describes the angle between the separation vector and the $x$ axis at some reference time. Note that the radiation pattern of the binary depends on $\iota$ and $\varphi_0$ (see, e.g., Eq.(\ref{eq:complex_h_from_hlm})).}
\label{fig:coord_frames}
\end{center}
\end{figure*}

The two polarizations $\hp(t)$ and $\hc(t)$ of GWs can be conveniently represented as a complex time-series $\h(t) \equiv \hp(t) - i\,\hc(t)$, which can be decomposed into spin $-2$ weighted spherical harmonic modes $\h_{\ell m}(t)$, so that the radiation along any direction $(\iota, \varphi_0)$ w.r.t. the source is given by  
\begin{equation}
\h(t; \iota, \varphi_0) = \sum_{\ell=2}^{\infty} \sum_{m=-\ell}^\ell \Ylm(\iota, \varphi_0) \, \hlm(t).
\label{eq:complex_h_from_hlm}
\end{equation}
Above, $\Ylm(\iota, \varphi_0)$ are the spin $-2$ weighted spherical harmonic basis functions where $\iota$ denotes the angle between the line-of-sight from the detector to the source and the total angular momentum of the binary, and $\varphi_0$ denotes the initial phase angle of the binary (see Fig.~\ref{fig:coord_frames}). The waveform $h(t)$ observed at the detector is a linear combination of the two polarizations $\hp(t)$ and $\hc(t)$:
\begin{equation}
h(t-t_0) = \frac{1}{d_L} \, \Big[\Fp(\theta, \phi, \psi)\,\hp(t) + \Fc(\theta, \phi, \psi)\,\hc(t)\Big],
\label{eq:hoft_det}
\end{equation}
where $d_L$ is the luminosity distance to the source, $t_0$ is the time of arrival of the signal at the detector, and $\Fp(\theta, \phi, \psi)$ and $\Fc(\theta, \phi, \psi)$ are the antenna pattern functions of the detector:  
\begin{eqnarray}
F_+ &=& \frac{1}{2}(1+\cos^2 \theta)\,\cos 2\phi\,\cos 2 \psi -
\cos \theta\,\sin 2\phi\,\sin2\psi \,, \nonumber \\
F_\times &=& \frac{1}{2}(1+\cos^2 \theta)\,\cos 2\phi\,\sin 2 \psi +
\cos \theta\,\sin 2\phi\,\cos2\psi\,.\nonumber \\
\label{eq:antennaPatterns}
\end{eqnarray}
Angles $\theta$ and $\phi$ denote the polar and azimuth angles of the binary on the sky measured in the detector frame, and $\psi$ is the polarization angle (see Fig.~\ref{fig:coord_frames}). The signal observed in a detector depends on the following set of parameters (assuming that the compact objects have negligible spin angular momenta): $\blambda = \{m_1, m_2, t_0, \varphi_0, \theta, \phi, \iota, \psi, d_L\}$.

GW signals $h(t)$ from binary black holes, buried in the background noise $n(t)$, are extracted using the technique of {matched filtering}, which is the optimal filtering to extract signals of known shapes buried in stationary Gaussian noise. {Matched filtering} involves maximizing the correlation of the data $d(t) \equiv h(t) + n(t)$ with a (normalized) template waveform $\hat{x}$. This provides a detection statistic, the signal-to-noise ratio (SNR), which is maximized over a ``bank'' of templates corresponding to different parameters:  

%
\begin{equation}
\label{eq:snr}
\rho = \max\limits_{\blambda} ~ \Big \langle d,~\hat{x}~(\blambda) \Big\rangle,
\end{equation}
where the angular brackets denote the following inner product of two time series $a(t)$ and $b(t)$
\be
\label{eq:correltation}
\Big\langle a,b \Big\rangle \equiv 4 \, \mathrm{Re} \int_{f_0}^{\infty} \frac{ \tilde{a}(f) \, \tilde{b}^*(f) }{ S_n(f) } df.
\ee
Above, $S_n(f)$ is the one-sided power spectral density (PSD) of the noise $n(t)$, $\tilde{a}(f)$ denotes the Fourier transform of $a(t)$, and a $^*$ indicates complex conjugation. The lower cutoff frequency $f_0$ is determined by the seismic wall of the detector noise. The normalized template waveforms is defined as $\hat{x} \equiv x/\sqrt{\Big\langle x,x \Big\rangle}$.

If the detector noise is well approximated by a stationary Gaussian process, a threshold on the SNR $\rho$ can be used to claim a detection corresponding to a certain false alarm probability. The optimal SNR $\rhoopt$ in detecting a signal is achieved when the template exactly matches with the signal. Thus,
\begin{equation}
\label{eq:snr_optimal}
\rhoopt^2 = \left<h,~{h}\right>.
\end{equation}
However, in an actual search it is unlikely that the template bank will contain a template waveform that matches exactly with the signal in the data. This can be due to the inaccuracies in modelling the template waveforms, discreteness of the template bank, etc. Thus, the SNR obtained is suboptimal:
\begin{equation}
\label{eq:snr_suboptimal}
\rhosubopt = \rhoopt ~.~ \FF,
\end{equation}
where $\FF$ is called the \emph{fitting factor}~\cite{Apostolatos:1995pj}, defined as:
\begin{equation}
\label{eq:fitting_factor}
\FF \equiv \max \limits_{\blambda} \left< \hat{h} , \hat{x}\,(\blambda) \right>.
\end{equation}
Thus, the fitting factor describes the fraction of optimal SNR that can be obtained using a suboptimal template family/bank, and is thus a useful quantity in characterizing the \emph{effectualness}~\cite{DIS98} of a template family/bank $x(\blambda)$ in detecting a target signal $h$. Note that, for a fixed SNR threshold, $\FF$ is directly related to the ``distance reach'' of a search, and $\FF^3$ to the ``volume reach''.

It is evident [see, e.g., Eqs.~(\ref{eq:hoft_det}), (\ref{eq:snr_optimal}) and (\ref{eq:snr_suboptimal})] that the distance/volume reach is a function of not only the \emph{intrinsic} parameters ($m_1, m_2$) of the binary, but also some of the \emph{extrinsic} parameters ($\theta,\phi,\iota,\psi,\varphi_0$). For example the SNR, and hence the distance/volume reach is the largest towards ``face-on'' ($\iota = 0, \pi$) binaries and the lowest for ``edge-on'' ($\iota = \pi/2$) binaries. It is useful to define the \emph{effective volume} of a search, defined as the fraction of the volume reach by an optimal search, averaged over the angles $\theta,\phi,\iota,\psi,\varphi_0$ after choosing appropriate distributions for these angles: 
\begin{equation}
\label{eq:efftive_vol}
V_\mathrm{eff}\,(m_1,m_2) = \frac{\overline{\rhosubopt^3}}{\overline{\rhoopt^3}},
\end{equation}
where the bars indicate averages over $\theta,\phi,\iota,\psi,\varphi_0$. We can also define the \emph{effective fitting factor} $\FFe$, defined as the cube root of the effective volume
\begin{equation}
\label{eq:effective_FF}
\FFe \,(m_1,m_2) = V_\mathrm{eff}\,(m_1,m_2)^{1/3}.
\end{equation}
If a template family has $\FFe \geq 0.965$, this means that the (average) loss of search volume due to the mismatch between the template family and the actual signal is less than $\sim10\%$. In this paper, we will use $\FFe = 0.965$ as a benchmark for deciding the effectualness of a template family.

If we interpret the parameter set $\blambda_\mathrm{max}$ that maximizes the inner product in~Eq.~(\ref{eq:fitting_factor}) as the parameters of the binary, which can be in general different from the true parameters $\blambda_\mathrm{true}$, this will result in the following systematic bias in the estimated parameters:
\begin{equation}
\label{eq:systerr}
\Delta \blambda = |{\blambda_\mathrm{max} - \blambda_\mathrm{true}}|,
\end{equation}
where $|~~|$ denotes the absolute value.

Similar to the $\FF$ and SNR, the systematic biases also depend on the parameters $\blambda$. We would like to use a single number (similar to $\FFe$) that quantifies the average bias in estimating the parameters of the binaries that are detectable. For this purpose we use the $\rhosubopt^3$ weighted average of the systematic biases and call it the \emph{effective bias}.

\be
\label{eq:avg_syst_err}
\Delta \blambda_\mathrm{eff}(m_1,m_2) = \frac{ \overline{ \Delta \blambda ~.~ \rhosubopt^3 } } {\overline{ \rhosubopt^3 } },
\ee
where the bars indicate averages over $\theta,\phi,\iota,\psi,\varphi_0$. We use $\rhosubopt^3$ as the weighting factor as it is proportional to the volume accessible to the search using quadrupole templates and is therefore proportional to the number of detectable sources.

GW measurements, like any other measurement in the presence of noise, will also have an associated statistical error. In the limit of high SNR, one reasonable way of estimating the expected statistical error (see, e.g.,~\cite{Vallisneri:2007ev} for caveats) is by using the Cramer-Rao inequality: the error covariance matrix $C_{\alpha\beta}$ is given by 
\begin{equation}
C_{\alpha\beta} \geq \Gamma_{\alpha\beta}^{-1}~,
\end{equation}
where $\Gamma_{\alpha\beta}$ is the Fisher information matrix:
\begin{equation}
\Gamma_{\alpha\beta} = \left<\partial_\alpha x,~ \partial_\beta x\right>.
\end{equation}
Above, $\partial_\alpha x$ denotes the partial derivative of the waveform ${x}(f)$ with respect to the parameter $\lambda_\alpha$, and the angle brackets denote the inner products defined in Eq.~(\ref{eq:correltation}). The rms error in measuring the parameter $\lambda_\alpha$ is $\sigma_\alpha = C_{\alpha\alpha}^{1/2}$. A template family can be considered \emph{faithful}~\cite{DIS98} to the signal if the systematic bias is considerably smaller than the expected statistical error. In this paper, we will take $(\Delta \blambda_\mathrm{eff})_\alpha \leq \sigma_\alpha$ as the benchmark for the faithfulness of a template family.

\begin{figure*}[htb]
\begin{center}
\includegraphics[width=7in]{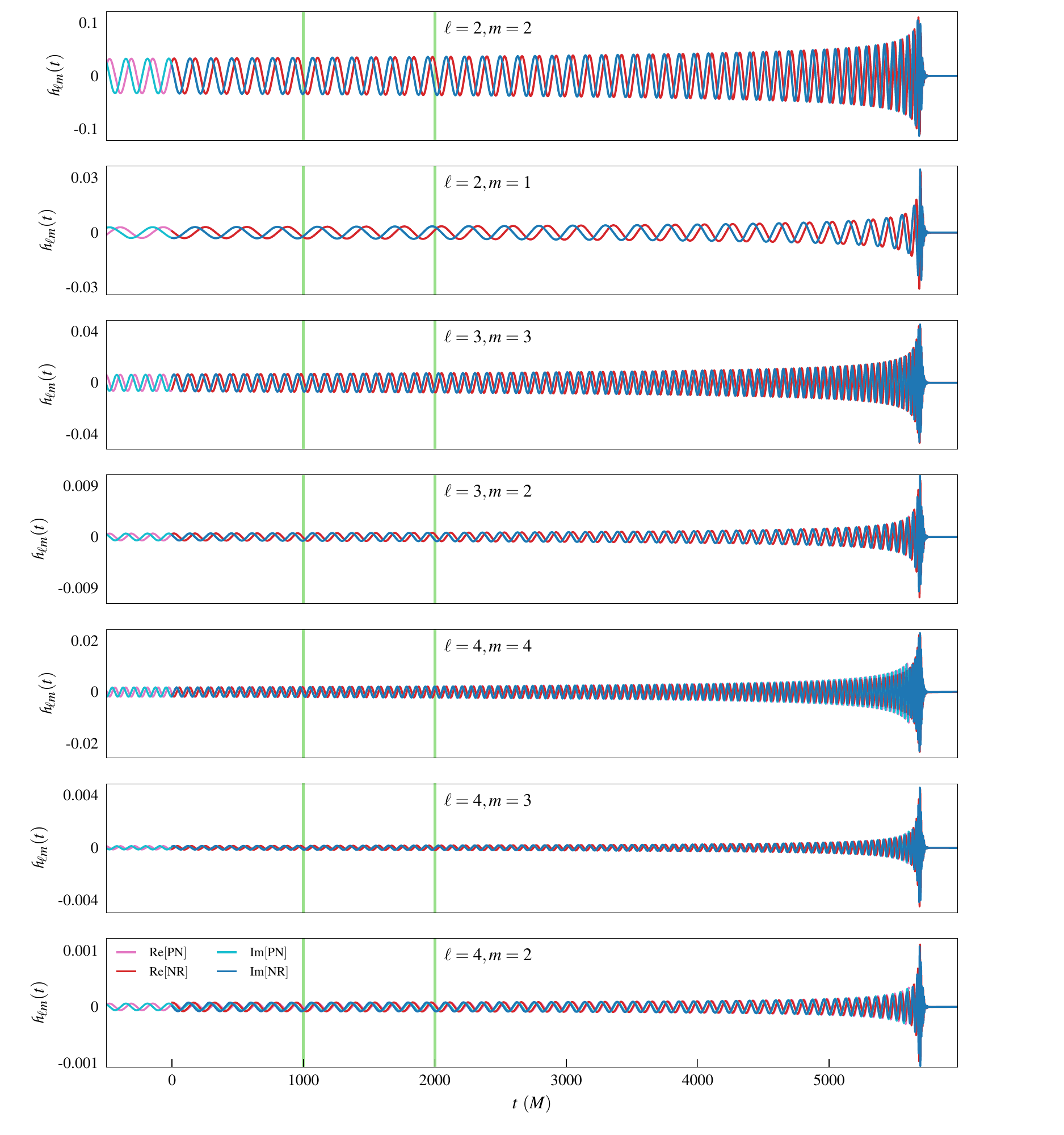}
\caption{Example of hybrid waveform modes constructed by matching NR and PN modes. These hybrid waveforms are constructed by matching non-spinning, $q=8$ NR waveforms computed using the SpEC code with PN/EOB waveforms describing the early inspiral. The horizontal axes show the time (with origin at the start of the NR waveforms) and the vertical axes show the GW modes $\hlm(t)$. The matching region $(1000 M, 2000 M)$ is marked by vertical green lines.}
\label{fig:hybrid_modes}
\end{center}
\end{figure*}

\section{Methodology}
\label{sec:methods}

\subsection{Numerical-relativity simulations}

We use two sets of NR waveforms: For mass ratio $q\leq 8$ we use waveforms computed by the SpEC code~\cite{Ossokine:2013zga,Hemberger:2012jz,Szilagyi:2009qz,Boyle:2009vi,Scheel:2008rj,Boyle:2007ft,Scheel:2006gg,Lindblom:2005qh,Pfeiffer:2002wt,SpEC,Mroue:2012kv,Mroue:2013xna, Buchman:2012dw}, kindly made public by the SXS collaboration~\cite{SXS-Catalog}. The SpEC code evolves conformally flat quasi-equilibrium initial data~\cite{Buchman:2012dw,Lovelace:2008tw,Pfeiffer:2007yz,Caudill:2006hw,Cook:2004kt} with the generalized harmonic formulation of general relativity~\cite{Friedrich:1996hq,Garfinkle:2001ni,Pretorius:2004jg}, using a pseudospectral multi-domain method for spatial discretization, and implements co-rotating coordinate system via the dual frame method~\cite{Scheel:2006gg}.

For mass ratio $q=18$, new NR simulations have been performed with the BAM code~\cite{Bruegmann:2006at,Husa:2007hp}. This code evolves black-hole-binary puncture initial data \cite{Brandt:1997tf,Bowen:1980yu} generated using a pseudo-spectral elliptic solver~\cite{Ansorg:2004ds}. Initial parameters for low-eccentricity inspiral were produced using integrations of the PN equations of motion, as described in~\cite{Husa:2007rh,Hannam:2010ec,Purrer:2012wy}. The numerical evolution is carried out with the $\chi$-variant of the moving-puncture \cite{Campanelli:2005dd,Baker:2005vv,Hannam:2006vv} version of the BSSN \cite{Shibata:1995we,Baumgarte:1998te} formulation of the 3+1 Einstein  evolution equations. Spatial finite-difference derivatives are sixth-order accurate in the bulk \cite{Husa:2007hp}, Kreiss-Oliger dissipation terms converge at fifth order, and a fourth-order Runge-Kutta algorithm is used for the time evolution. A grid hierarchy of 15 levels of refinement boxes is used, where the innermost cubic mesh refinement boxes are roughly a factor 1.5 larger than the black hole horizons, and correspond to $96^3$ grid-points (not counting buffer zones and a reduction by a factor of 2 by using manifest equatorial symmetry). The free function $\eta$ in the gamma-freezing shift condition (see Eq. (27) in \cite{Bruegmann:2006at}), which controls the size of the black holes,  is set to $\eta=1$. 

The GWs emitted by the binary are calculated from the Newman-Penrose scalar $\Psi_4$. For the SpEC waveforms, $\Psi_4$ was extrapolated to future null infinity, while in the case of the BAM waveform, we used the $\Psi_4$ extracted at the largest available extraction radius ($160M$). GW strain is computed from $\Psi_4$ using the fixed-frequency-integration algorithm described in~\cite{Reisswig:2010di}. Recent comparative discussion of the SpEC and BAM codes, together with other numerical codes used for evolving black hole binaries, are given in~\cite{Ajith:2012az,Hinder:2013oqa}. Parameters of the NR waveforms used in this paper are summarized in Table~\ref{tab:nr_waveforms}. 

\begin{table}
\centering
\begin{tabular}{c@{\quad} c@{\quad}c@{\quad}c@{\quad}r}
\toprule
Simulation ID & $q$ & $M\omega_\mathrm{orb}$ & $e$ & \# orbits \\
\midrule
SXS:BBH:0090 & $1$  &  $0.011$ &  $9.9 \times 10^{-4}$ &  $32.4$\\
SXS:BBH:0169 & $2$ &  $0.018$ &  $1.2 \times 10^{-4}$ &  $15.7$ \\
SXS:BBH:0167 & $4$ &  $0.021$ &  $9.9 \times 10^{-5}$ &  $15.6$\\
SXS:BBH:0166 & $6$ & $0.019$ &  $4.4 \times 10^{-5}$ &  $21.6$\\
SXS:BBH:0063 & $8$ &  $0.019$ &  $2.8 \times 10^{-4}$ &  $25.8$ \\
BAM:q18a0a0& $18$ & $0.041$ & $2.8 \times 10^{-3}$ & $6.6$ \\
\bottomrule
\end{tabular}
\caption{Summary of the parameters of the NR waveforms used in this paper: $q \equiv m_1/m_2$ is the mass ratio of the binary, $M \omega_\mathrm{orb}$ is the orbital frequency after the junk radiation and $e$ is the residual eccentricity.  }
\label{tab:nr_waveforms}
\end{table}

\subsection{Post-Newtonian inspiral waveforms}

The spherical harmonics modes (scaled to unit total mass and unit distance) of the PN inspiral waveforms, 3.5PN accurate in phase and 3PN accurate in amplitude can be written as  
\begin{equation}
\hlm^\mathrm{PN}(t) = 2 \eta v^2 \sqrt{\frac{16\pi}{5}} \, H_{\ell m} \, e^{-i\, m \varphi_\mathrm{orb}(t)},
\label{eq:pn_modes}
\end{equation}
where the mode amplitudes $H_{\ell m}$ are computed up to 3PN accuracy by~\cite{Blanchet:2008je} while the 3.5PN orbital phase $\varphi_\mathrm{orb}(t)$ can be computed in the adiabatic approximation using inputs given in \cite{Blanchet:2004ek} and references therein. 

However, we have found that, for higher mass ratios ($q \gtrsim 8$) the phase evolution predicted by the standard PN approximants differ appreciably from the template family (EOBNRv2) used in this study, during the late inspiral. Since EOBNRv2 is used as the template waveform, the mismatch due to the difference in phase evolution can be misinterpreted as an effect of non-quadrupole modes. In order to avoid this, we compute the phase evolution of the inspiral part from the $\ell = m = 2$ mode of the EOBNRv2 waveforms. That is,  
\begin{equation}
\hlm^\mathrm{PN}(t) = 2 \eta v^2 \sqrt{\frac{16\pi}{5}} \, H_{\ell m} \, e^{-i\, m \varphi_\mathrm{EOB22}(t)/2},
\label{eq:pn_modes_eobpn}
\end{equation}
where $\varphi_\mathrm{EOB22}$ is the phase of the $\ell = m = 2$ mode of the EOBNRv2 waveform. Note that, for $m = 2$ modes, $H_{\ell m}$ contains imaginary terms at order 2.5PN and above, which can be absorbed into the phase. However, since this correction appears at order 5PN and above in the phase, we neglect these corrections and use $|H_{\ell m}|$ instead of $H_{\ell m}$ for the $m = 2$ modes. 

\subsection{Construction of hybrid waveforms}
\label{sec:hybrid_waveforms}
We construct a set of \emph{hybrid} waveforms containing all the relevant modes by matching NR waveforms with PN waveforms with the same intrinsic binary parameters, using a generalization of the method introduced in~\cite{Ajith:2007qp}. Note that the frames with respect to which the NR and PN waveforms are decomposed into spherical harmonics modes can be different (see Sec.~\ref{sec:gw_intro}). These frames need to be aligned with each other before matching the NR modes $\hlm^\mathrm{NR}(t)$ with PN modes $\hlm^\mathrm{PN}(t)$. In general three Euler rotations ($\iota, \varphi_0, \psi$) can be performed between the two frames. However, one angle ($\iota$) is fixed by the choice of aligning the $z$ axis of both (PN and NR) frames along the direction of the total angular momentum of the binary, which is uniquely defined (while different conventions can be followed in defining the other two angles). Note that the two Euler angles $\varphi_0$ and $\psi$ can be absorbed into one if we are only considering one value of $m$, as in previous work on quadrupole modes. 

We match the PN modes with NR modes by a least square fit over two rotations ($\varphi_0, \psi$) on the NR waveform and the time-difference between NR and PN waveforms:
\begin{equation}
\delta = \mathrm{min}_{t_0,\varphi_0, \psi} \int_{t_1}^{t_2} dt \sum_{\ell,m} \left|\hlm^\mathrm{NR}(t-t_0) e^{i (m \varphi_0 + \psi)}  - \hlm^\mathrm{PN}(t) \, \right|.
\end{equation}
Note that $\delta$ represents the integrated difference between NR and PN waveforms over an an appropriately chosen matching interval $(t_1,t_2)$, where the NR and PN calculations are assumed to be accurate. The hybrid waveforms are constructed by combining the NR waveform with the ``best matched'' PN waveform in the following way:
\begin{equation}
\hlm^\mathrm{hyb}(t) \equiv \, \tau(t) \, \hlm^\mathrm{NR}(t-t_0') \ e^{i(m\varphi_0'+\psi')} + (1-\tau(t)) \, \hlm^\mathrm{PN}(t) ,
\end{equation}
where $t_0', \varphi_0'$ and $\psi'$ are the values of $t_0, \varphi_0$ and $\psi$ that minimizes the difference $\delta$ between PN and NR waveforms. Above, $\tau(t)$ is a weighting function defined by:
\begin{eqnarray}
\tau(t) \equiv \left\{ \begin{array}{ll}
0 & \textrm{if $t < t_1 $}\\
\frac{t-t_1}{t_2-t_1}  & \textrm{if $t_1 \leq t < t_2 $}\\
1 & \textrm{if $t_2 \leq t$.}
\end{array} \right.
\label{eq:HybWaveWeight}
\end{eqnarray}
For $q\leq8$, the matching region $(t_1,t_2)$ was chosen to be $(1000 M, 2000 M)$, where $t = 0$ is defined as the start time of ``clean'' NR data after the junk radiation. The orbital frequencies corresponding to the start and the end of the matching region range from $M \omega_\mathrm{orb_1} \in (0.012, 0.023)$ and $M \omega_\mathrm{orb_2} \in (0.012, 0.029)$, depending on the length of the NR waveform. The NR waveform was shorter for $q = 18$. Hence the matching region was chosen to be $(100 M, 400 M)$, corresponding to $M \omega_\mathrm{orb_1} = 0.042$ and $M \omega_\mathrm{orb_2} = 0.048$. 

We consider spherical harmonic modes up to $\ell = 4$ and $m = -\ell ~ \mathrm{to} ~ \ell$ in this analysis, except the $m = 0$ modes. An example of the hybrid waveform modes for a non-spinning binary with $q = 8$ is shown in Fig.~\ref{fig:hybrid_modes}. It can be seen that higher modes are excited only during the very late inspiral, merger and ringdown. The effect of higher modes will be appreciable only in the mass range where the SNR contributed by the merger-ringdown is a significant fraction of the total SNR. This is the reason we restrict our study to the mass range $20 M_\odot \leq M \leq 250 M_\odot$. 

\label{sec:effectualness}
\begin{figure*}[tbh]
\begin{center}
    \subfigure[$~q=1$, $M = 100 \Mo$\label{subfig:skymap_snr_q1_m100}]{\includegraphics[scale=\skymapscale]{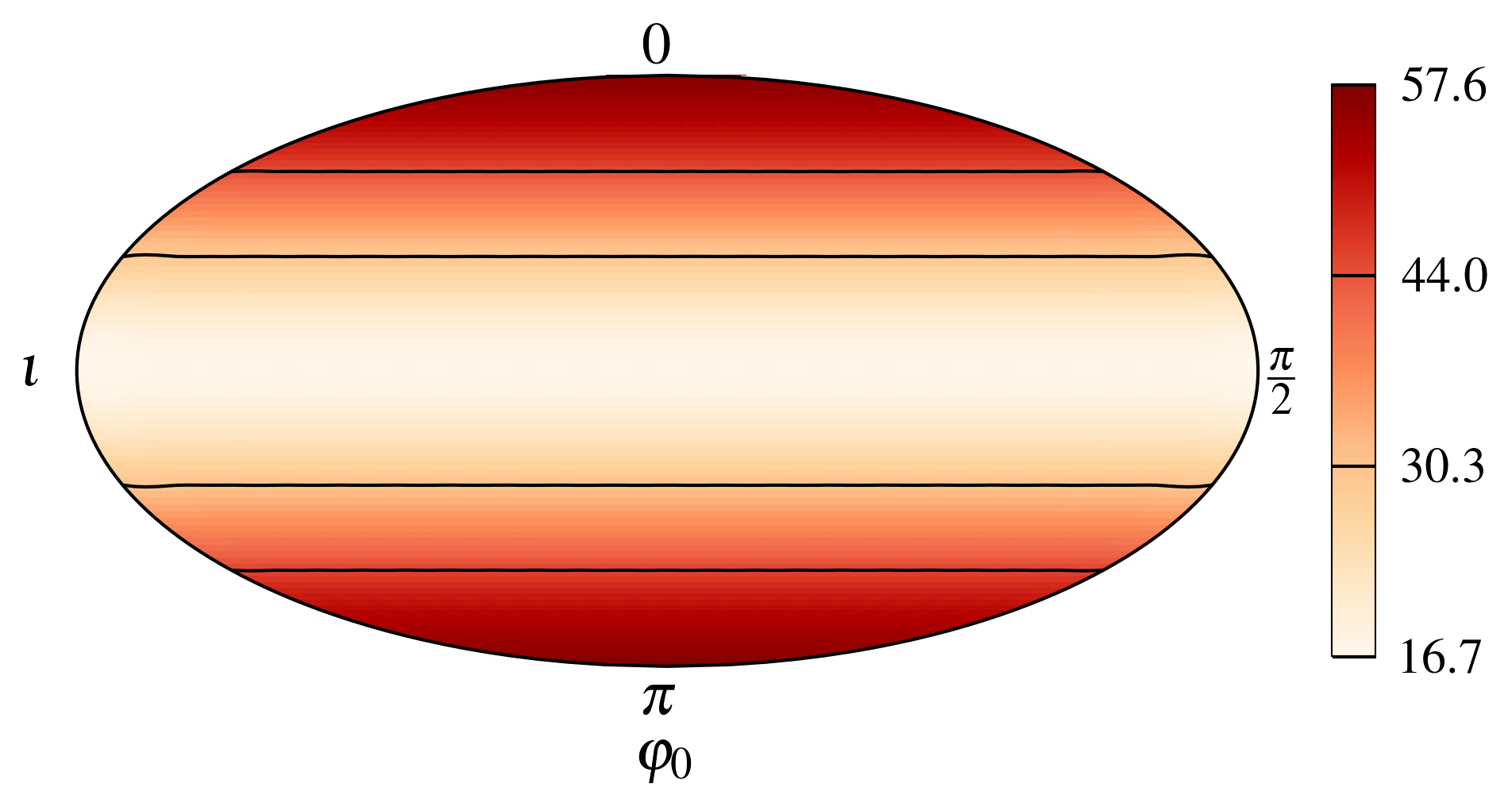}}
    \subfigure[$~q=8$, $M = 100 \Mo$\label{subfig:skymap_snr_q8_m100}]{\includegraphics[scale=\skymapscale]{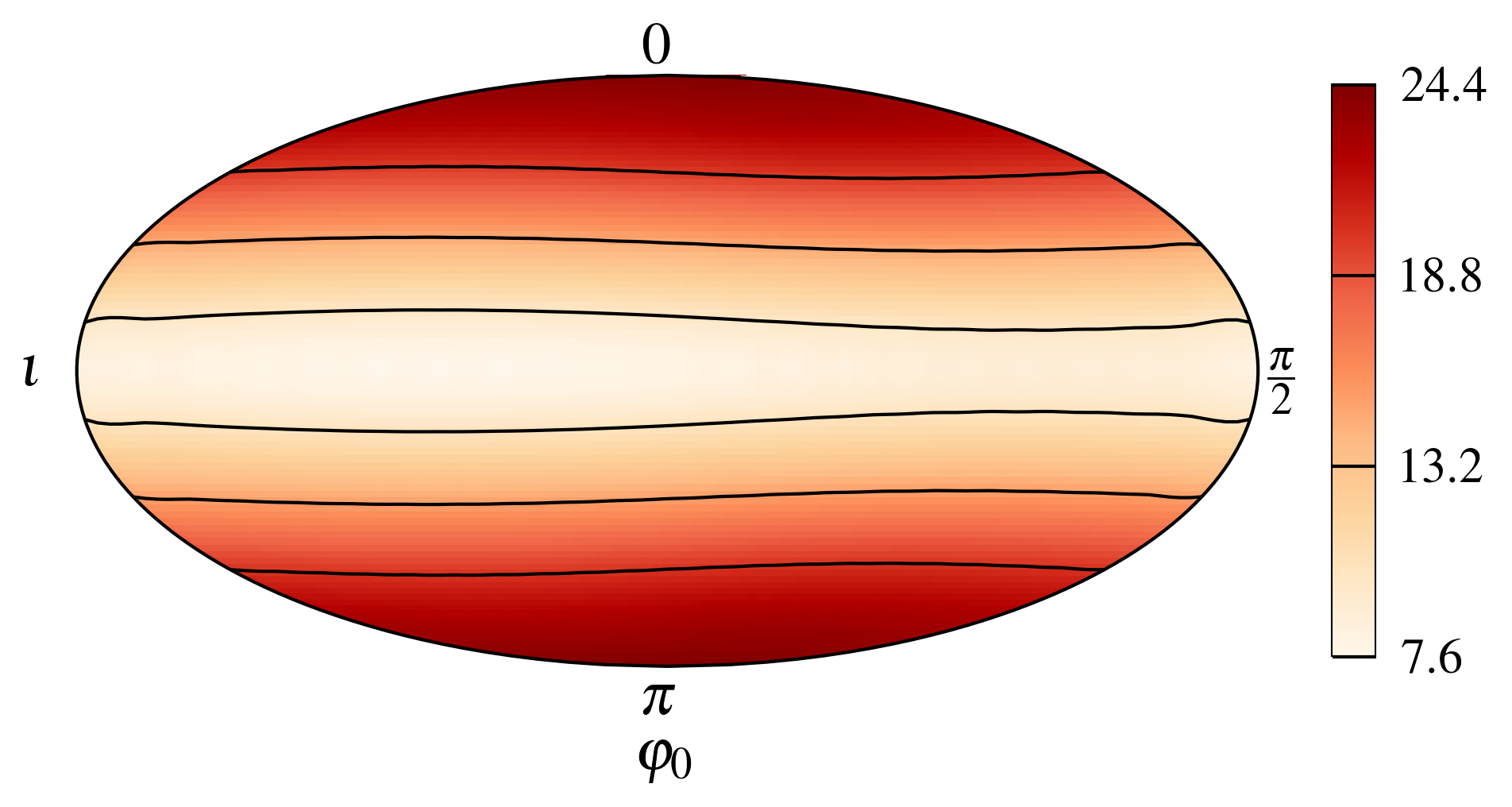}}
\caption{Optimal SNR averaged over polarization angle $\psi$ for binaries located at 1 Gpc. The y-axis shows the inclination angle $\iota$ in radians and the x-axis shows the initial phase of the binary $\varphi_0$ in radians. The left (right) corresponds to binaries with mass ratio $q = 1 \, (q = 8)$ and total mass $M = 100 M_\odot$.}
\label{fig:skymap_snr}
\end{center}
\end{figure*}

\begin{figure*}[tbh]
\begin{center}
    \subfigure[$~q=1$, $M = 100 \Mo$\label{subfig:skymap_ff_q1_m100}]{\includegraphics[scale=\skymapscale]{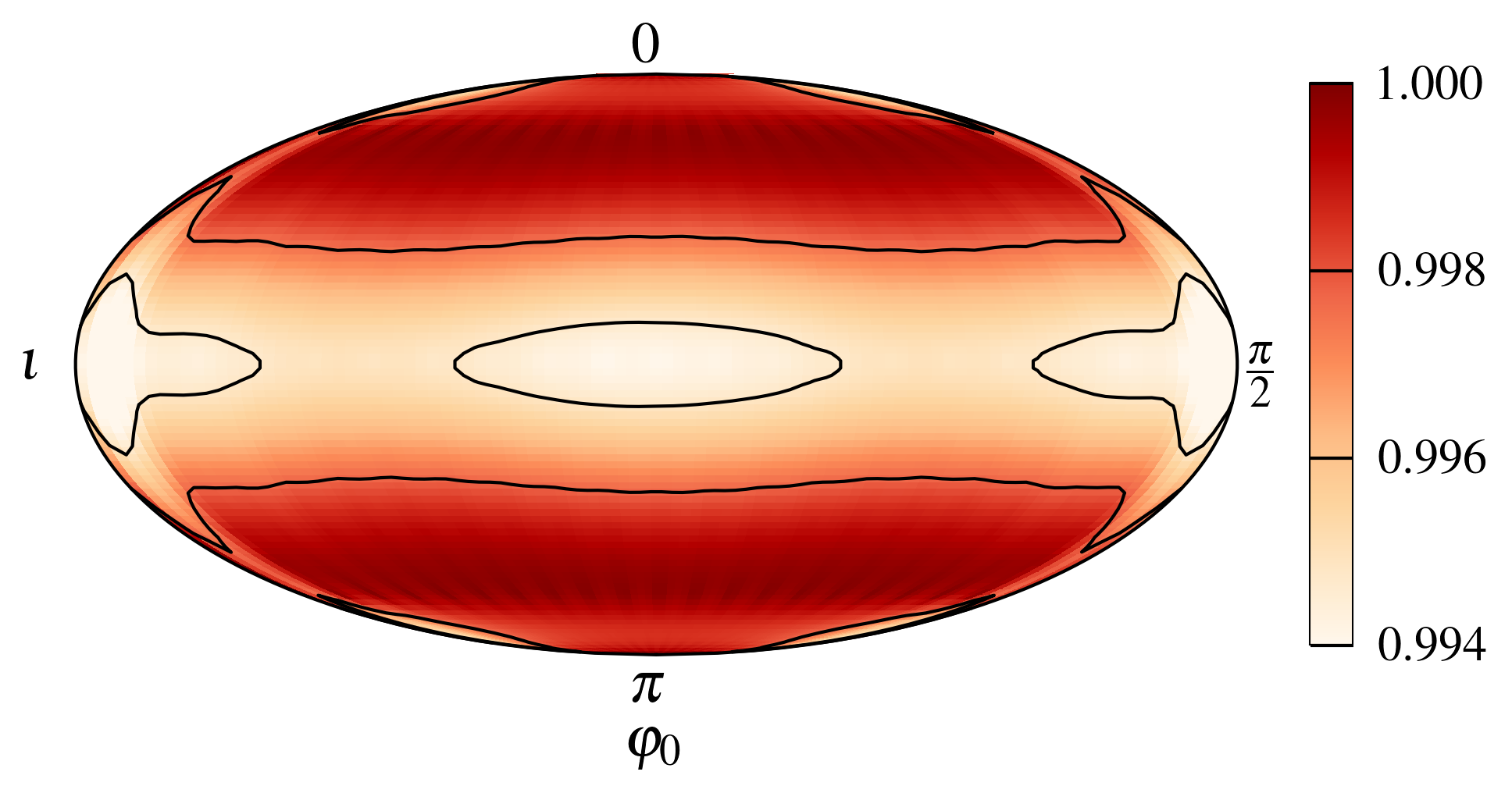}}\quad
    \subfigure[$~q=8$, $M = 100 \Mo$\label{subfig:skymap_ff_q8_m100}]{\includegraphics[scale=\skymapscale]{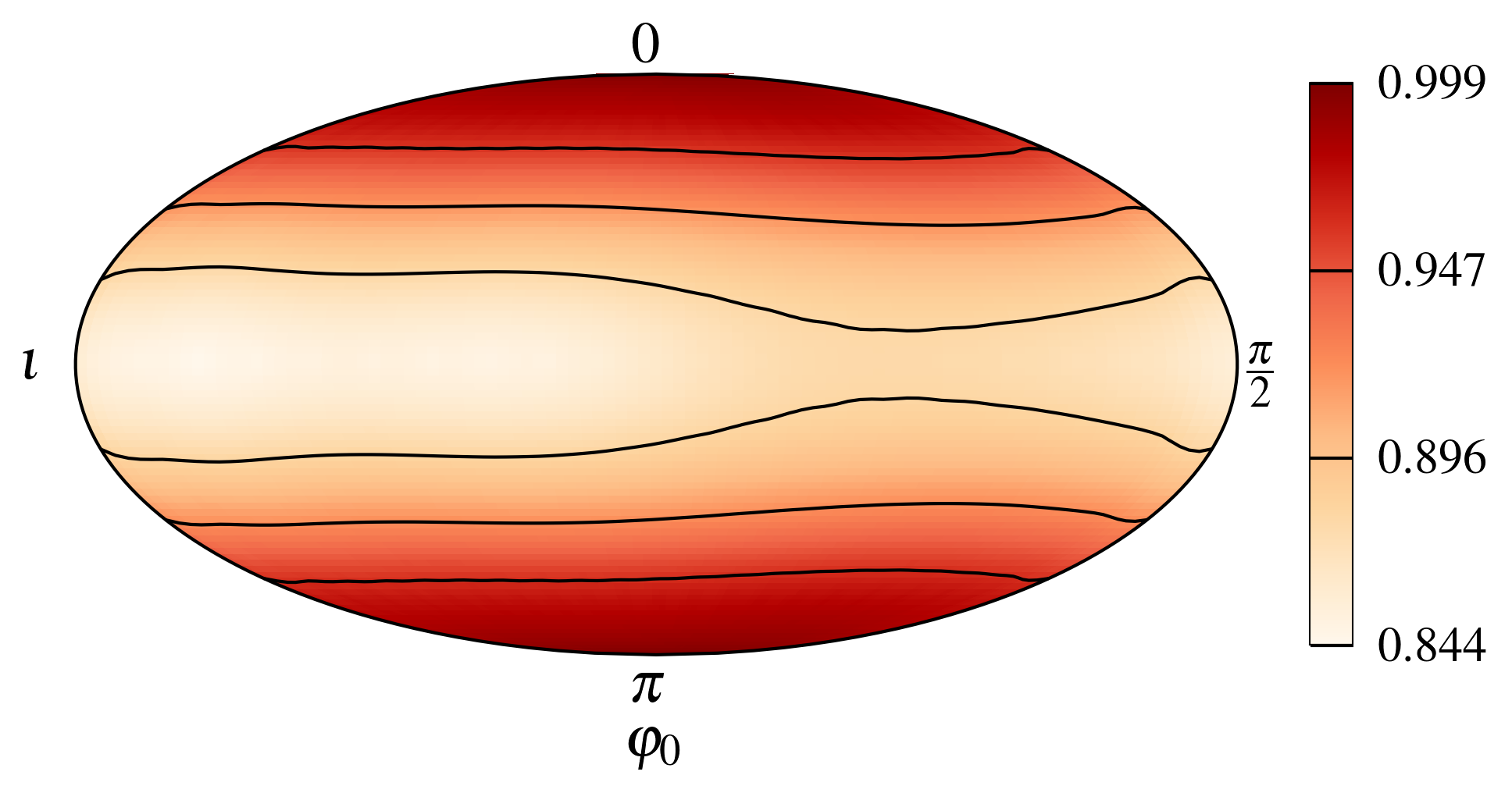}}\quad
\caption{Fitting factor of quadrupole templates for different orientation angles, averaged over polarization angle $\psi$. The y-axis shows the inclination angle $\iota$ in radians and the x-axis shows the initial phase of the binary $\varphi_0$ in radians. The left (right) panel correspond to binaries with mass ratio $q = 1 \, (q = 8)$ and $M = 100 M_\odot$. It may be noted that the fitting factor is smallest (largest) at $\iota = \pi/2 ~(\iota = 0,\pi)$ where contribution from the non-quadrupolar modes is the largest (smallest).}
\label{fig:skymap_ff}
\end{center}
\end{figure*}

\subsection{Choice of template waveforms}
\label{sec:templates}

We use the quadrupole modes  ($\ell = 2, m = \pm2$ modes) of the EOBNRv2~\cite{Pan:2011gk} waveform family as detection templates for this study. These waveforms have very good agreement with the quadrupole modes of the hybrid waveforms discussed in the previous section. Note the EOBNRv2 also includes the effect of non-quadrupole modes. However, since this study aims to understand the effect of neglecting the non-quadrupole modes, we take only the quadrupole modes of EOBNRv2 as templates. The waveforms are generated in time-domain using the LALSimulation~\cite{lalsimulation} software package. 

\subsection{Detector model, computation of the fitting factor}
\label{sec:ff_computation}
In our study we use the ``zero-detuned, high-power" design noise PSD~\cite{adligo-psd} of Advanced LIGO with a low frequency cut-off of 20 Hz. To compute the fitting factor [see Eq.~(\ref{eq:fitting_factor})], the maximization of the inner product over the two template parameters $\varphi_0$ and $t_0$ is performed using the standard techniques -- by taking the absolute value of the inner product defined in Eq.~(\ref{eq:correltation}) and by maximizing the correlation function by means of a Fast Fourier Transform. Maximization of the inner product over the mass parameters is performed using the Nelder-Mead down-hill simplex maximization algorithm as implemented in SciPy~\cite{scipy}. We choose to do this maximization in the two dimensional space of chirp mass ${\mathcal {M}} \equiv M\eta^{\frac{3}{5}}$ and symmetric mass ratio $\eta \equiv m_1 m_2 / M^2$. 

\section{Results and discussion}
\label{sec:results}

\subsection{Effectualness of quadrupole-mode templates}

\begin{figure}[tbh]
\begin{center}
\includegraphics[scale=0.455]{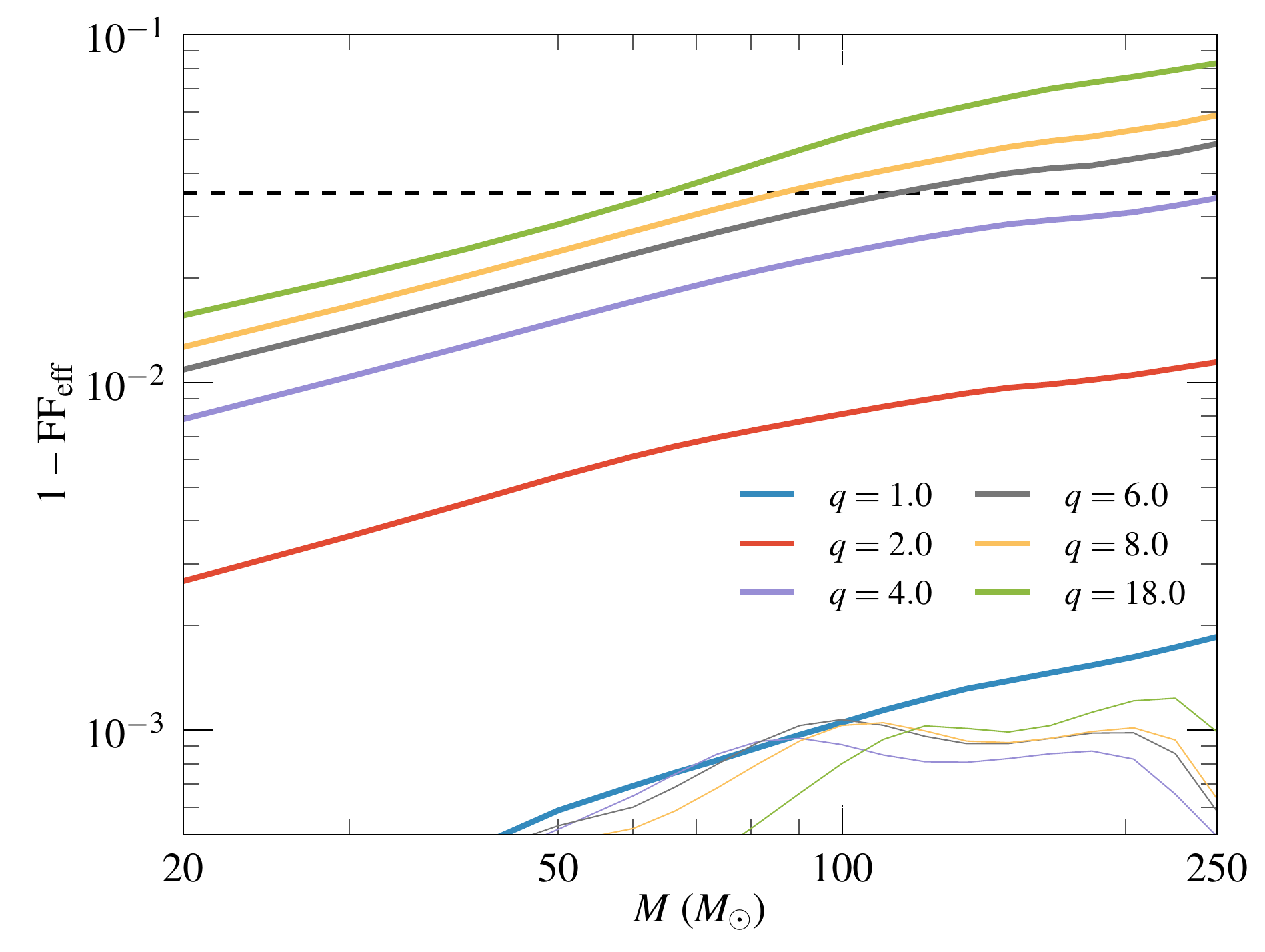}
\caption{Thick lines show the ``ineffectualness'' (1 - $\FFe$) of quadrupole mode templates towards hybrid waveforms including sub-dominant modes, while the thin lines show the same towards hybrid waveforms including only the quadrupole ($\ell=2,m=\pm2$) modes. The horizontal axis reports the total mass of the binary while the mass ratio is shown in the legend. The horizontal dashed black line corresponds to $1-\FFe^3 = 10\%$. Note that some of the thin lines are not visible in this plot as their values are $\ll 10^{-3}$.}
\label{fig:effFF}
\end{center}
\end{figure}

In this section, we evaluate the {effectualness} of the quadrupole-mode templates by computing the fitting factor of a quadrupole-mode-only inspiral-merger-ringdown template family, EOBNRv2 against the hybrid waveforms described in Section~\ref{sec:hybrid_waveforms}. 

It is evident from Eqs.~(\ref{eq:complex_h_from_hlm}) and (\ref{eq:hoft_det}) that the observed GW signal $h(t)$ depends on angles $\iota$, $\varphi_0$, $\psi$, $\theta$ and $\phi$. However, the dependence of $h(t)$ on $\theta$ and $\phi$ comes as an amplitude scaling and a constant phase shift (see, e.g.,~\cite{Pekowsky:2013hm}). While the observed SNR has a strong dependence on $\theta$ and $\phi$, since the match between the signal and template is computed using normalized waveforms, the match has only very weak dependence on these angles. Hence we set $\theta=\phi=0$ in this study. The error introduced by this restriction is very small ($\sim 0.1\%$) due the weak dependence of the matches on $\theta, \phi$ and the strong selection bias towards binaries with $\theta \simeq 0, \pi$ (where the antenna pattern function peaks). 

Fig.~\ref{fig:skymap_snr} shows the optimal SNR of the hybrid waveforms at different values of $\iota$ and $\varphi_0$ (averaged over the polarization angle $\psi$). We see that the SNR is the largest for ``face-on'' orientations ($\iota = 0, \pi$; poles in the plots) and smallest for ``edge-on'' orientations ($\iota = \pi/2$; equator in the plots). This is due to the fact that contribution from the quadrupole modes (which are the dominant modes) are the largest for face-on orientations and the smallest for edge-on orientations. It can be seen from the right plot of Fig.~\ref{fig:skymap_snr} (which corresponds to a $q = 8$ binary) that the SNR drops less as $\iota \rightarrow \pi/2$, as compared to the left plot (which corresponds to a $q=1$ binary). This is a reflection of the fact that the contribution from sub-dominant modes increases with increasing mass ratio.

Figure~\ref{fig:skymap_ff} shows the $\FF$ of the EOBNRv2 templates towards hybrid waveforms constructed at different values of $\iota$ and $\varphi_0$ (averaged over the polarization angle $\psi$). It is clear that for the case of the equal-mass binary (left panels) there is practically no loss of the SNR for all orientations of the binary, while for the binary with mass ratio 8 (right panels), the $\FF$ can be as low as $\sim 0.84$ for binaries that are highly inclined with the detector. Note that the $\FF$ is still high near the face-on orientations and low near the edge-on orientation. This is explained by fact that the face-on orientation is almost entirely comprised of quadrupole mode, and the template is a good representation of the true signal at this orientation. In contrast, the relative contribution from the sub-dominant modes is the highest for the edge-on case, resulting in low FFs.

From these results we see that the orientations that are modeled least (most) faithfully by the quadrupole mode are also the orientations that have the least (most) luminosity, therefore mitigating the effect of sub-dominant modes and inherently reducing their importance, as noted by previous studies~\cite{Pekowsky:2013hm,Brown:2013hm,Capano:2013hm}.

As the $\FF$ varies significantly with different orientations, we evaluate the $\FF$ at all possible orientations of the binary with respect to the detector by varying $\cos \iota, \varphi_0$ and $\psi$ uniformly in $[-1,1],[0,2\pi)$ and $[0,2\pi)$ respectively. We then compute the {effective fitting factor} $\FFe$ by doing a weighted average of the FF values as defined in Eq.~(\ref{eq:effective_FF}). Figure~\ref{fig:effFF} shows $1-\FFe$ as a function of the total mass of the binary for different mass ratios. The thick lines show the ``{ineffectualness}'' of quadrupole mode templates towards ``full'' hybrid waveforms, and the corresponding thin lines show the same towards ``quadrupole-only'' hybrid waveforms. The difference between the two cases indicates the effect of sub-dominant modes on the detection problem. 

From the thick lines we see that the {ineffectualness} increases with increasing mass ratio, due to the fact that higher order modes are excited by a larger extent for binaries with high mass ratios. Also note the trend that the ineffectualness increases with the total mass of the binary. The sub-dominant modes are excited more prominently during the merger and ringdown stages of the coalescence and are therefore more important for high-mass binaries, for which the observed signal is dominated by the merger and ringdown. We set $\FFe \geq 0.965$ (which corresponds to a $\sim10\%$ loss in detection volume) as the benchmark for the relative importance of non-quadrupole modes in the detection. We see that $\FFe>0.965$ for binaries with $q\leq4$. However for higher mass ratios ($q > 4$) the {effective fitting factor} falls below $0.965$ for ``high-mass'' binaries. Figure~\ref{fig:summary_fig} summarizes the region in the parameter space where the loss of detection rate due to neglecting non-quadrupole modes is 
greater than 10\%.

\subsection{Systematic errors in estimating parameters}
\label{sec:syst_err}

\begin{figure*}[htb]
\begin{center}
    \subfigure[$~q=1$, $M = 100 \Mo$\label{subfig:skymap_mbias_q1_m100}]{\includegraphics[scale=\skymapscale]{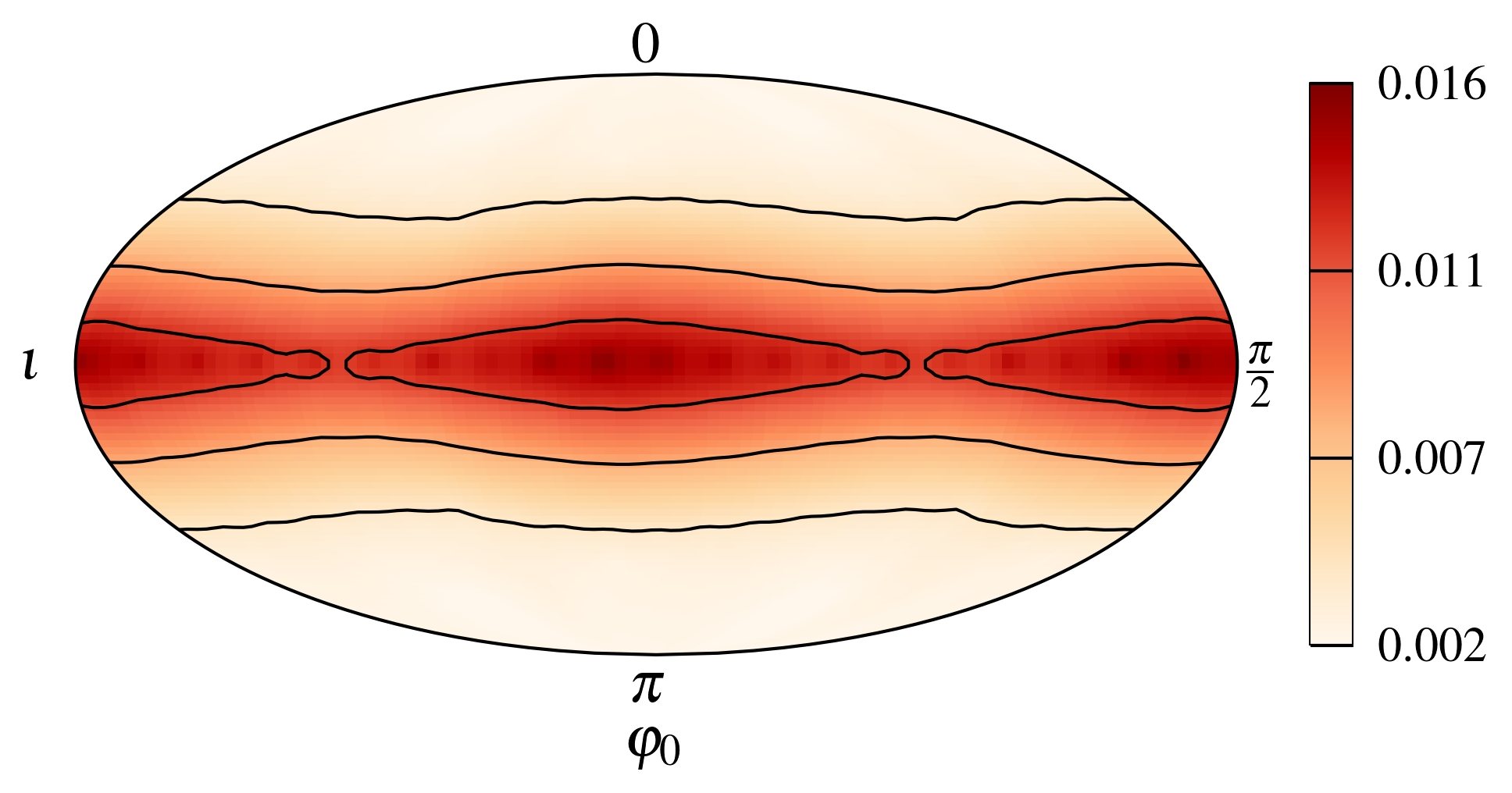}}
    \subfigure[$~q=8$, $M = 100 \Mo$\label{subfig:skymap_mbias_q8_m100}]{\includegraphics[scale=\skymapscale]{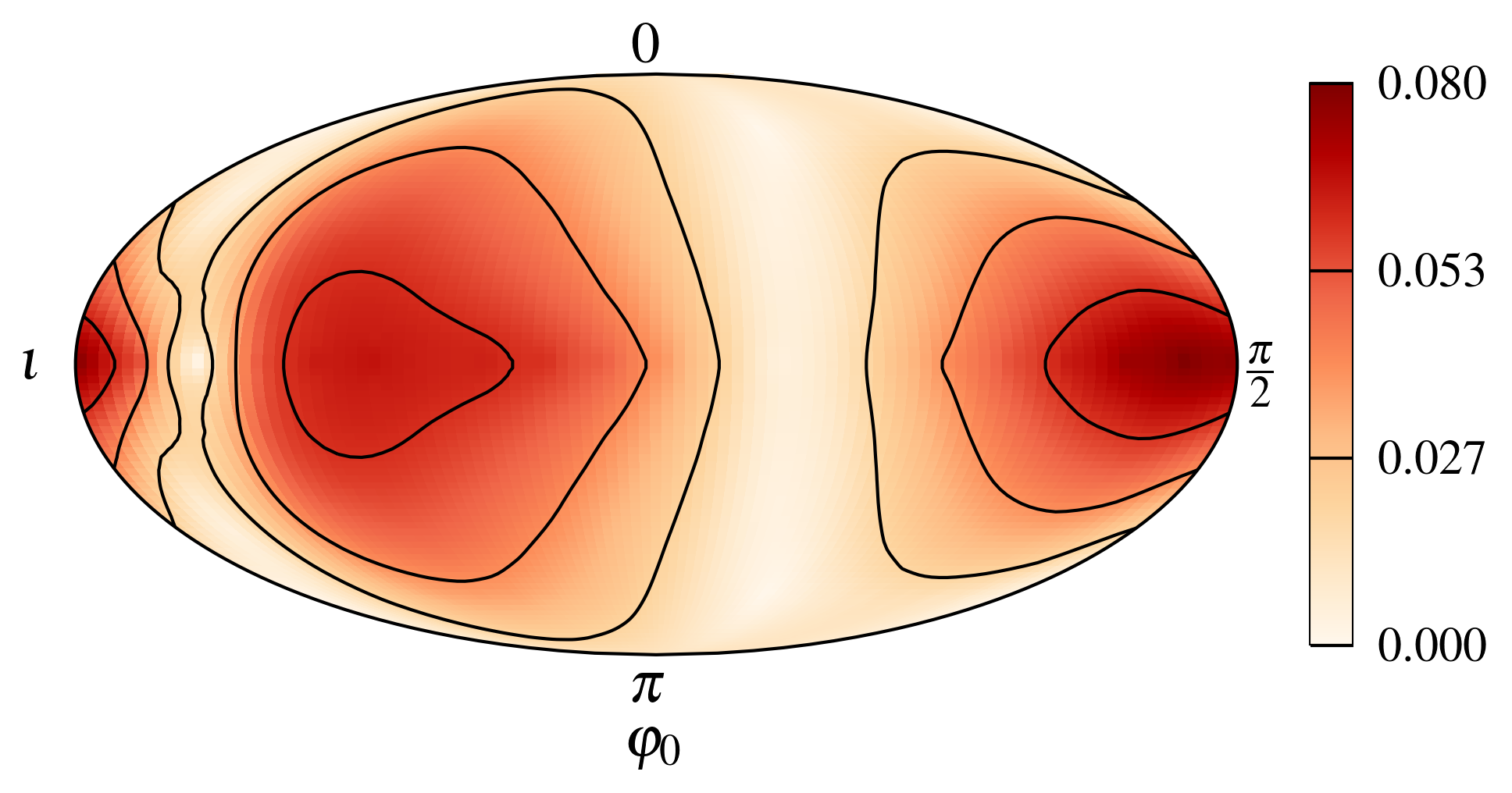}}
\caption{Systematic bias in the estimation of total mass $\Delta M/M$ averaged over polarization angle $\psi$. The y-axis shows the inclination angle $\iota$ in radians and the x-axis shows the initial phase of the binary $\varphi_0$ in radians.}
\label{fig:skymap_mbias}
\end{center}
\end{figure*}

In this section we study the systematic errors in the estimated parameters (total mass $M$ and symmetric mass ratio $\eta$) of BBHs due to neglecting non-quadrupole modes. We evaluate the fractional systematic biases at all possible orientations of the binary with respect to the detector after varying $\cos \iota, \varphi_0$ and $\psi$ uniformly in $[-1,1],[0,2\pi)$ and $[0,2\pi)$ respectively. As an example, the relative systematic bias in estimating the total mass $M$ for different values of $\iota$ and $\varphi_0$ (averaged over the polarization angle $\psi$) is shown in Fig.~\ref{fig:skymap_mbias}. 

\begin{figure*}[htb]
\begin{center}
        \includegraphics[scale=\paramestscale]{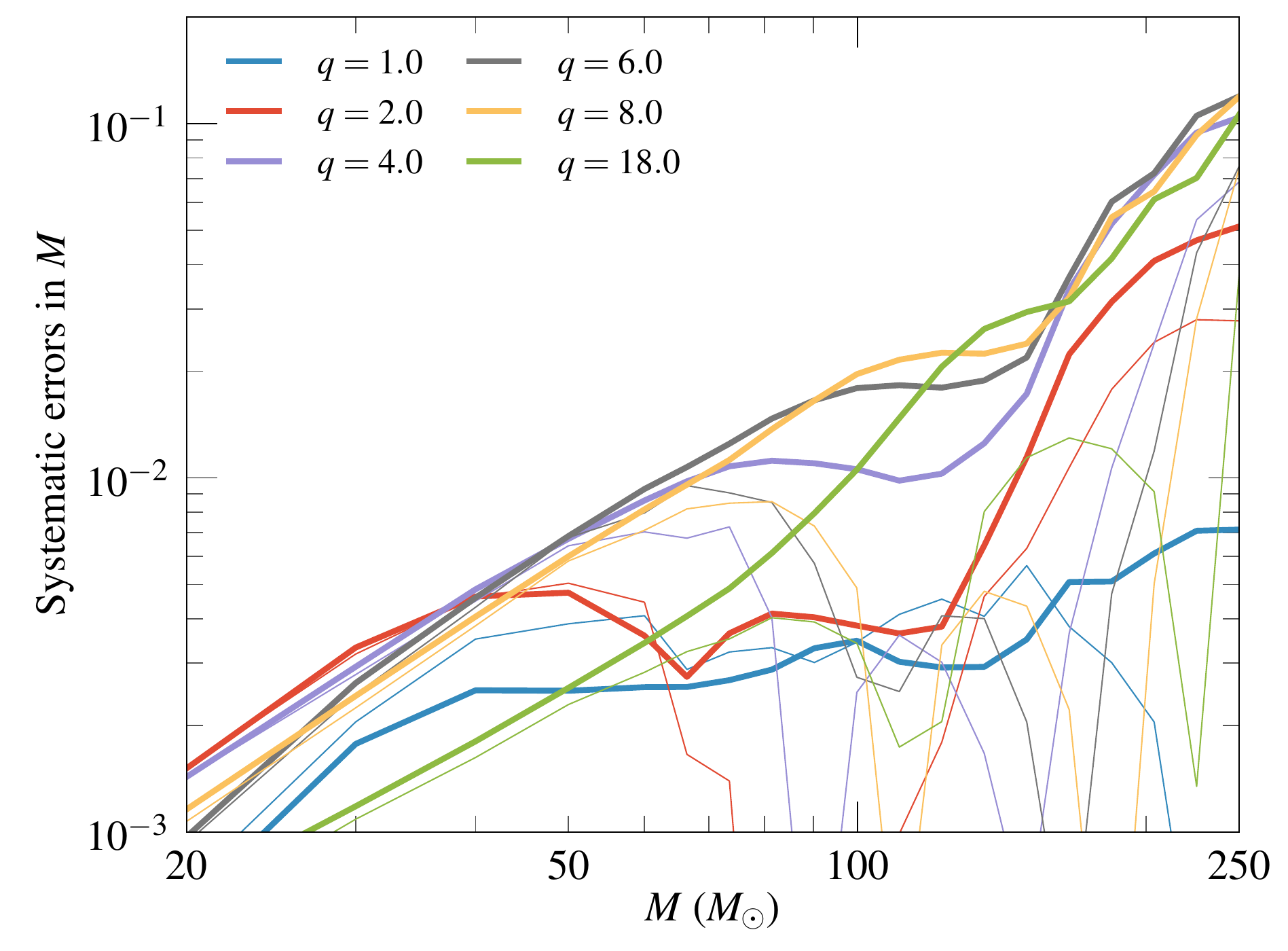}
        \includegraphics[scale=\paramestscale]{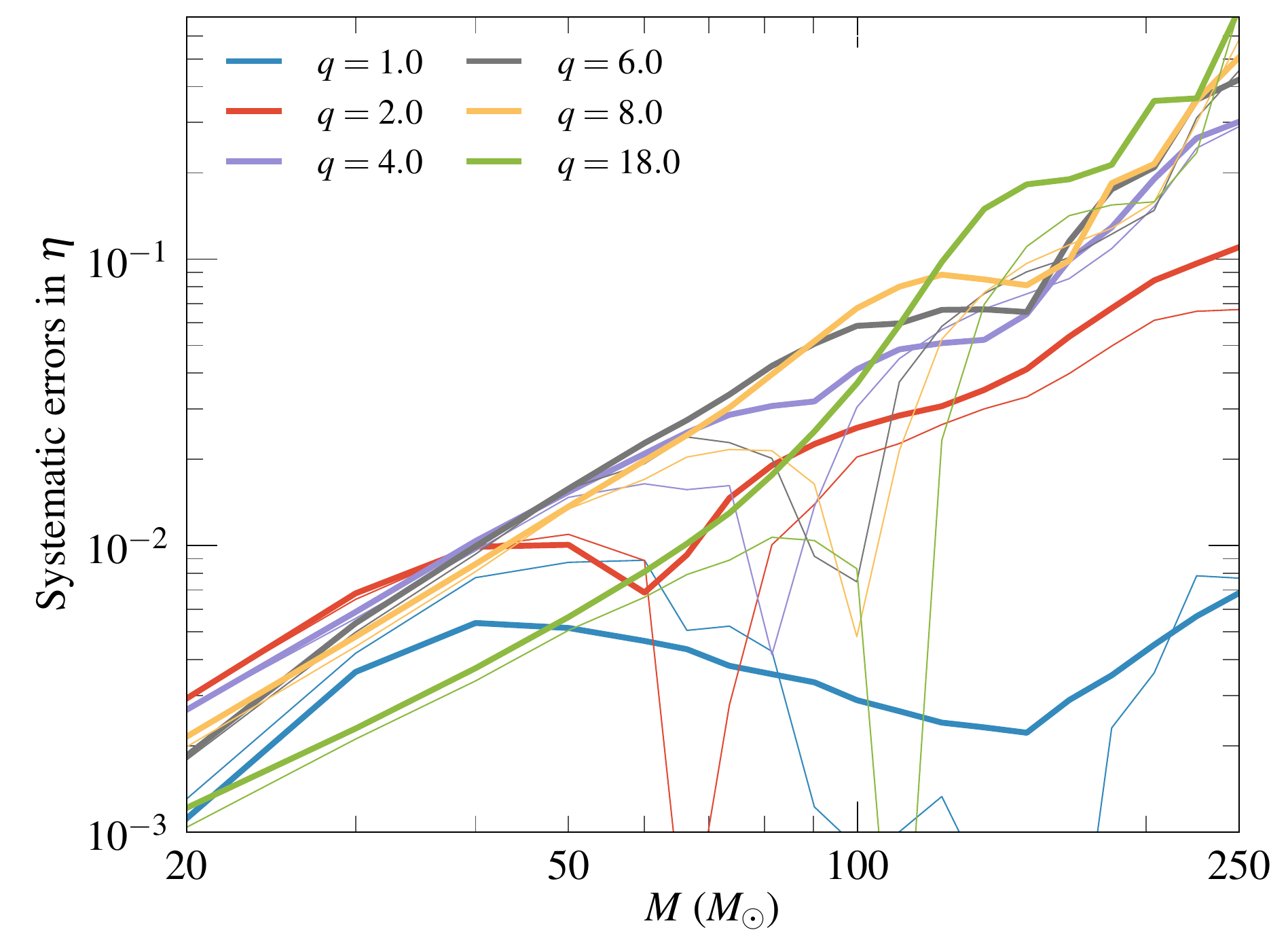}
\caption{The effective bias (fractional) in estimating the parameters total mass $M$ (left) and symmetric mass ratio $\eta$ (right) using quadrupole mode templates. The thick lines correspond to the errors assuming that ``full'' hybrid waveforms as the target signals, while the thin lines correspond to the errors assuming that ``quadrupole-only'' hybrid waveforms as the target signals. The systematic errors in estimating $M$ are generally dominated by the errors in neglecting non-quadrupole modes, for binaries with $q > 1$ and $M \gtrsim 70 M_\odot$. On the other hand, the systematic errors in estimating $\eta$ are dominated by the same effect only in a small, intermediate mass range $(70 M_\odot \lesssim M \lesssim 120 M_\odot )$.}
\label{fig:param_syserr}
\end{center}
\end{figure*}

The \emph{effective bias} [see Eq.~(\ref{eq:avg_syst_err})] in estimating the parameters $M$ and $\eta$ as a function of the total mass of the binary for different mass ratios is plotted in Fig.~\ref{fig:param_syserr}. As before, the thick lines correspond to the systematic errors assuming that the target signals are ``full'' hybrid waveforms. The corresponding thin lines show the systematic errors assuming that target waveforms are ``quadrupole'' hybrid waveforms (i.e., the systematic errors due to the inaccurate modeling of the quadrupole modes). The difference between the two cases gives an indication of the systematic errors introduced due to neglecting the non-quadrupole modes in the templates. If the solid lines are well above the corresponding thin lines, this indicates that the error budget is dominated by the effect of non-quadrupole modes. The systematic errors in estimating $M$ are generally dominated by the errors in neglecting non-quadrupole modes, for binaries with $q > 1$ and $M \gtrsim 70 M_\odot$. On the other hand, the systematic errors in estimating $\eta$ are dominated by the same effect only in a small, intermediate mass range $(70 M_\odot \lesssim M \lesssim 120 M_\odot )$.


\begin{figure*}[htb]
\begin{center}
    \includegraphics[scale=\paramestscale]{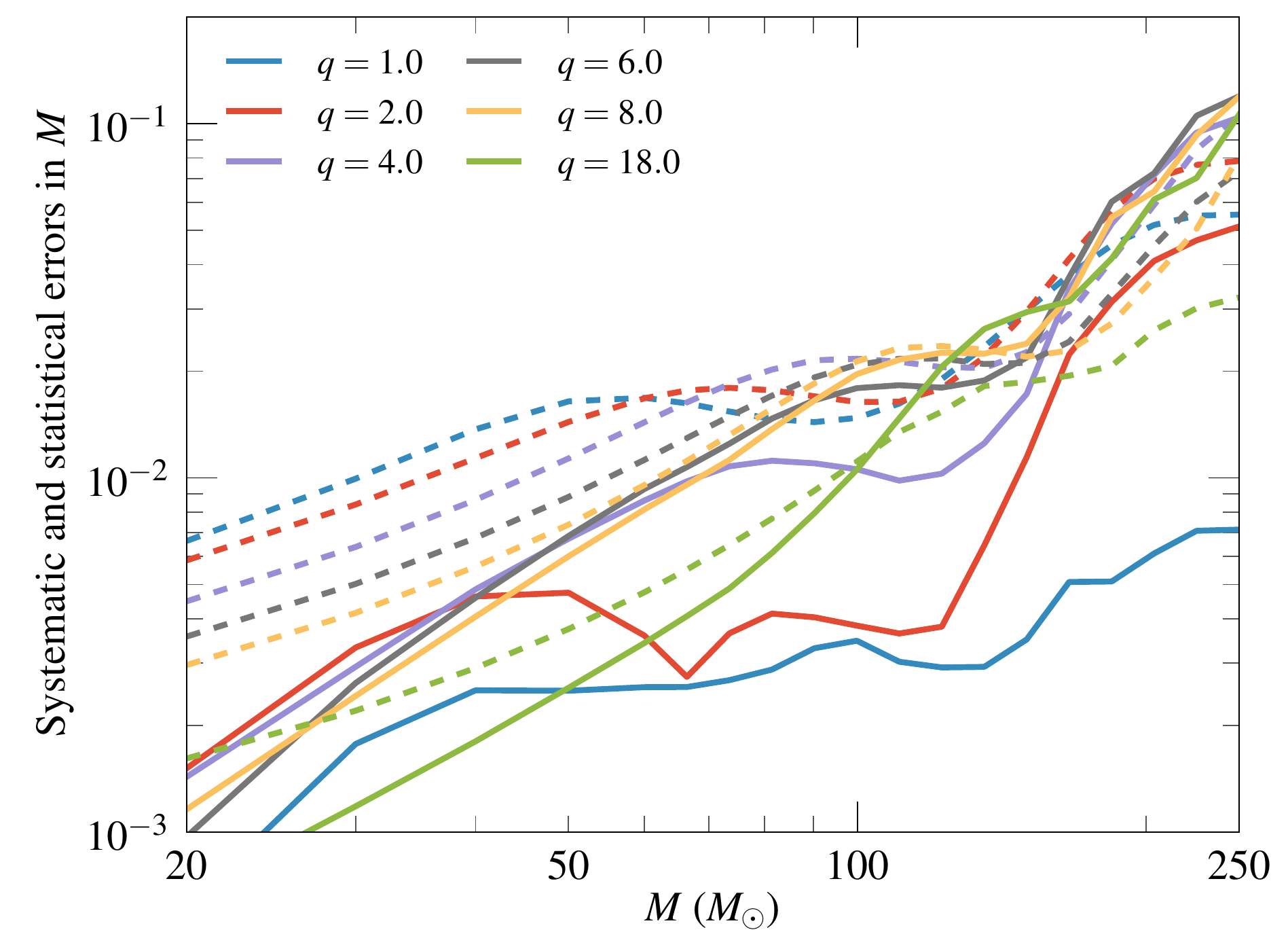}
    \includegraphics[scale=\paramestscale]{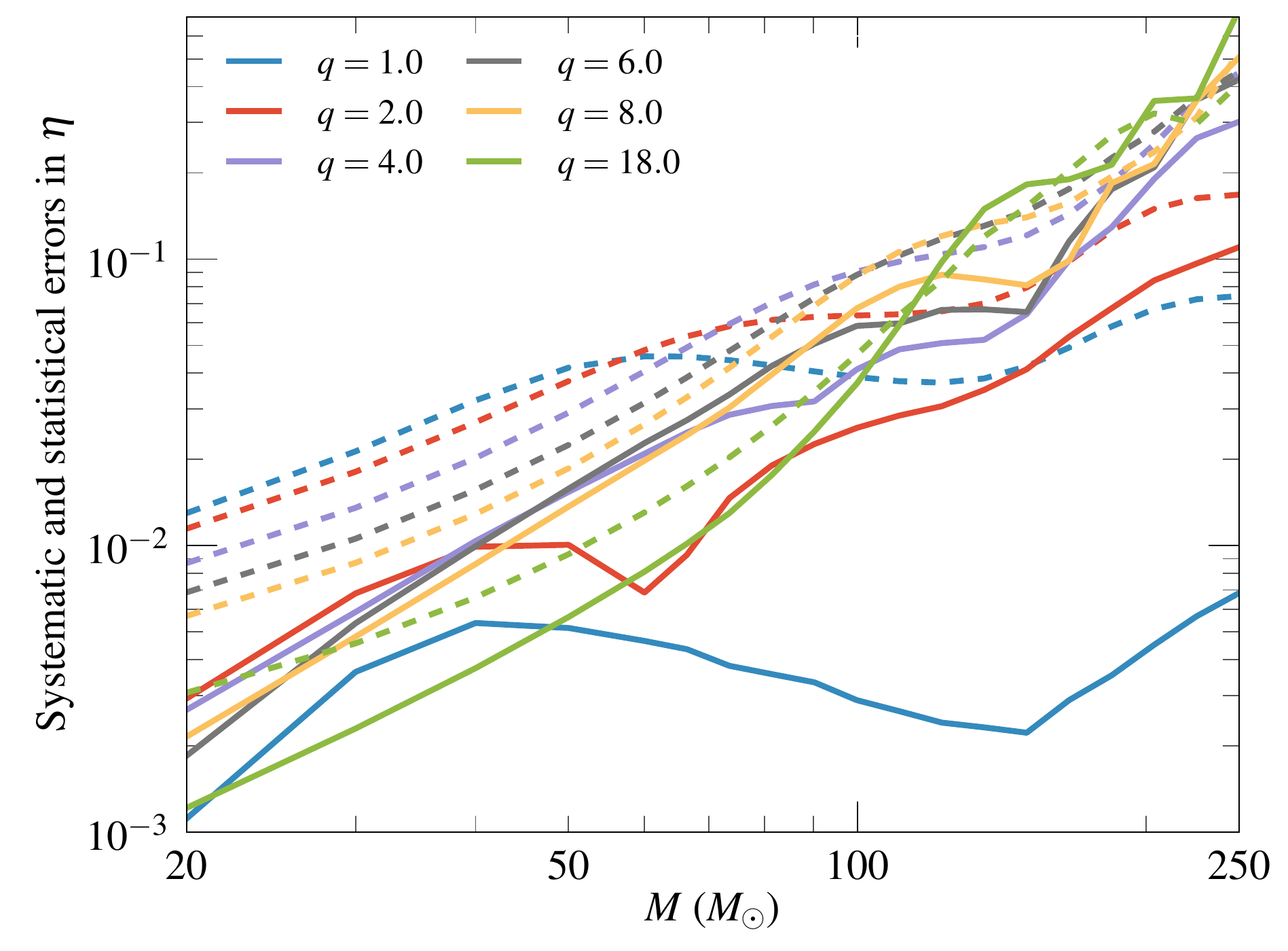}
\caption{The solid lines correspond to effective bias (fractional) in estimating the parameters total mass $M$ (left) and symmetric mass ratio $\eta$ (right) using quadrupole mode templates, assuming that ``full'' hybrid waveforms as the target signals. The dashed lines correspond to the statistical errors (fractional) in estimating the same parameters. In the computation of the statistical errors, we assume that the binaries are observed with SNR of 8 (averaged over the sky-location and orientation of the binary).}
\label{fig:param_staterr}
\end{center}
\end{figure*}

Let us note that, as long the systematic errors are significantly lower than the statistical errors, it is safe to ignore the systematic errors. Statistical errors are fundamental limits to a measurement due to the intrinsic stochasticity of the noise. In order to gauge the relative importance of the systematic errors discussed above, we compare them against the expected statistical errors from a search using quadrupole templates. The statistic errors are evaluated using a Fisher matrix analysis, taking the sources at a constant SNR of 8 averaged over all angles ($\theta, \phi, \iota, \psi, \varphi_0$). Figure~\ref{fig:param_staterr} compares the $1\,\sigma$ statistical errors (dashed lines) in estimating $M$ and $\eta$ using quadrupole mode templates with the effective systematic bias (solid lines) in parameter estimation of the same assuming that target waveforms contain all the relevant modes. It can be seen that the error budget in the parameter estimation of $M$ is, in general, dominated by systematic errors for high-mass ($M \gtrsim 150 M_\odot$) binaries with large mass ratio ($q \gtrsim 4$), while the estimation of $\eta$ is in dominated by the statistical errors over almost the entire parameter space under consideration. Figure~\ref{fig:summary_fig} summarizes the region in the parameter space where the error budget is dominated by the systematic errors. 

The fact that there is a region in the parameter space (bottom left region in Fig.~\ref{fig:summary_fig}) where non-quadrupole modes are important for detection, but not for parameter estimation may seem surprising. A closer look at Figs.~\ref{fig:skymap_snr}, \ref{fig:skymap_ff}, \ref{fig:skymap_mbias} will reveal the cause: For the case of highly unequal-mass binaries, a search using quadrupole-only templates preferentially selects binaries with face-on orientation (due to the low fitting factor of quadrupole-only templates towards highly inclined binaries). Among the observed binaries the contribution from non-quadrupole modes is negligible, and hence they make little impact on parameter estimation. There is also a region in the parameter space (top right region in Fig.~\ref{fig:summary_fig}) where non-quadrupole modes are important for parameter estimation, but not for detection. In this high-mass region, due to the small number of cycles in the detector band, quadrupole-mode templates are able to mimic the full-mode signal at the cost of introducing a large systematic bias in the estimated total mass. 

\section{Conclusion}
We studied the effects of sub-dominant modes in the detection and parameter estimation of non-spinning BBHs using advanced GW detectors.  As target signals we used hybrid waveforms constructed by matching NR simulations describing the late inspiral, merger and ringdown of the coalescence with PN/EOB waveforms describing the early inspiral. These signals contained contributions from all modes up to $\ell = 4$ and $m = -\ell ~ \mathrm{to} ~ \ell$ except the $m=0$ modes. Our study considered non-spinning BH binaries with total masses $20\Mo \leq M \leq 250\Mo$, mass ratios $1 \leq q \leq 18$ and all angles describing the orientation of the binary. We quantified the effect of non-quadrupole modes on detection in terms of the effective fitting factor (cube root of the fractional detection volume) and the effect on parameter estimation in terms of the effective bias in the estimated parameters. Although several of these aspects have been studied in the past, we believe that this paper provides a comprehensive summary of the effect of non-quadrupole modes in the detection and parameter estimation of binary black holes. Figure~\ref{fig:summary_fig} shows the regions in the parameter space where the contribution from non-quadrupole modes is important for GW detection and parameter estimation. 

Let us also list the limitations of this work. While our study was restricted to the case of non-spinning BBHs, we expect the searches and parameter estimation in Advanced LIGO/Virgo data to employ spinning waveform models, most likely aligned-spin models for searches, and generic-spinning models for parameter estimation~\cite{lvc-whitepaper-2014-2015}. It is unclear how our conclusions will change in the presence of spins. The precision and accuracy with which the mass ratio can be measured is severely diminished by a partial degeneracy with the spin components parallel to the orbital angular momentum~\cite{Cutler:1994ys,Poisson:1995ef,Baird:2012cu,Hannam:2013uu}, but this can be mitigated somewhat when parameter estimation is performed with a generic spinning waveform model~\cite{Chatziioannou:2014coa} and also when the binary's orientation makes precession effects detectable~\cite{O'Shaughnessy:2014dka}. However, since the main contributor to higher modes is the mass ratio (the dominant modes in precessing systems are still confined to $\ell = 2$), we expect our broad conclusions to continue to hold. Also, while we studied the loss of SNR due to neglecting non-quadrupole modes, we did not study their effect on signal-based vetoes such as the ``chi-square'' veto. Note that we estimated the expected statistical errors using the Fisher matrix formalism. Since the errors given by the Cram\'er-Rao bound are lower limits, our estimates on the region of the parameter space where the systematic errors due to neglecting non-quadrupole modes are negligible should be treated as conservative estimates. 

Employing search templates including the effect of non-quadrupole modes is likely to improve the detection rates of BBHs in certain regions in the parameter space. However, in order to quantify this we need to consider the possible increase in the false alarm rate due to the change in the distribution of the ``background'' (noise-generated triggers) when the detection statistic is maximized over additional parameters describing the relative orientation of the binary (see, e.g., Appendix A of ~\cite{Capano:2013hm}). In addition, we note that employing ``full-mode'' templates in parameter estimation is likely to reduce not only the systematic errors but also the statistical errors (due to the increased information content in the waveform). We leave some of these investigations as future work. 

\smallskip 
\acknowledgments 
We are indebted to the SXS collaboration for making a public catalog of numerical-relativity waveforms. Numerical simulations with the BAM code were performed on Mare Nostrum at the Barcelona Supercomputing Center and at SuperMUC, LRZ trough a European PRACE grant, and the Cardiff ARCCA cluster, while the data-analysis calculations were performed on the Mowgli cluster at ICTS-TIFR and the LDG cluster at IUCAA. We thank Evan Ochsner for useful comments on the manuscript and K. G. Arun, Luc Blanchet, Ashok Choudhary, Archisman Ghosh, Bala Iyer, Amruta Jaodand, Chandra Kant Mishra, and Harald Pfeiffer for useful discussions. PA's research was supported by a Ramanujan Fellowship from the Department of Science and Technology, India the SERB FastTrack fellowship SR/FTP/PS-191/2012, and by the AIRBUS Group Corporate Foundation through a chair in ``Mathematics of Complex Systems'' at ICTS-TIFR. JCB and SH were supported by the Spanish MIMECO grants FPA2010-16495 and CSD2009-00064, European Union FEDER funds, and  Conselleria d'Economia i Competitivitat del Govern de les Illes Balears. MH was supported by STFC grants ST/H008438/1 and ST/I001085/1, and MP by ST/I001085/1.

\bibliography{HigherModes}

\begin{thebibliography}{71}
\expandafter\ifx\csname natexlab\endcsname\relax\def\natexlab#1{#1}\fi
\expandafter\ifx\csname bibnamefont\endcsname\relax
  \def\bibnamefont#1{#1}\fi
\expandafter\ifx\csname bibfnamefont\endcsname\relax
  \def\bibfnamefont#1{#1}\fi
\expandafter\ifx\csname citenamefont\endcsname\relax
  \def\citenamefont#1{#1}\fi
\expandafter\ifx\csname url\endcsname\relax
  \def\url#1{\texttt{#1}}\fi
\expandafter\ifx\csname urlprefix\endcsname\relax\def\urlprefix{URL }\fi
\providecommand{\bibinfo}[2]{#2}
\providecommand{\eprint}[2][]{\url{#2}}

\bibitem[{\citenamefont{Damour et~al.}(1998)\citenamefont{Damour, Iyer, and
  Sathyaprakash}}]{DIS98}
\bibinfo{author}{\bibfnamefont{T.}~\bibnamefont{Damour}},
  \bibinfo{author}{\bibfnamefont{B.~R.} \bibnamefont{Iyer}}, \bibnamefont{and}
  \bibinfo{author}{\bibfnamefont{B.~S.} \bibnamefont{Sathyaprakash}},
  \bibinfo{journal}{Phys. Rev. D} \textbf{\bibinfo{volume}{57}},
  \bibinfo{pages}{885} (\bibinfo{year}{1998}).

\bibitem[{\citenamefont{Abadie et~al.}(2011)}]{Abadie:2011kd}
\bibinfo{author}{\bibfnamefont{J.}~\bibnamefont{Abadie}} \bibnamefont{et~al.}
  (\bibinfo{collaboration}{LIGO Scientific Collaboration, Virgo
  Collaboration}), \bibinfo{journal}{Phys.Rev.} \textbf{\bibinfo{volume}{D83}},
  \bibinfo{pages}{122005} (\bibinfo{year}{2011}), \eprint{1102.3781}.

\bibitem[{\citenamefont{Aasi et~al.}(2013)}]{Aasi:2012rja}
\bibinfo{author}{\bibfnamefont{J.}~\bibnamefont{Aasi}} \bibnamefont{et~al.}
  (\bibinfo{collaboration}{LIGO Scientific Collaboration, Virgo
  Collaboration}), \bibinfo{journal}{Phys.Rev.} \textbf{\bibinfo{volume}{D87}},
  \bibinfo{pages}{022002} (\bibinfo{year}{2013}), \eprint{1209.6533}.

\bibitem[{\citenamefont{Buonanno et~al.}(2007)}]{Buonanno:2007pf}
\bibinfo{author}{\bibfnamefont{A.}~\bibnamefont{Buonanno}}
  \bibnamefont{et~al.}, \bibinfo{journal}{Phys. Rev.}
  \textbf{\bibinfo{volume}{D76}}, \bibinfo{pages}{104049}
  (\bibinfo{year}{2007}), \eprint{0706.3732}.

\bibitem[{\citenamefont{Pan et~al.}(2011)\citenamefont{Pan, Buonanno, Boyle,
  Buchman, Kidder et~al.}}]{Pan:2011gk}
\bibinfo{author}{\bibfnamefont{Y.}~\bibnamefont{Pan}},
  \bibinfo{author}{\bibfnamefont{A.}~\bibnamefont{Buonanno}},
  \bibinfo{author}{\bibfnamefont{M.}~\bibnamefont{Boyle}},
  \bibinfo{author}{\bibfnamefont{L.~T.} \bibnamefont{Buchman}},
  \bibinfo{author}{\bibfnamefont{L.~E.} \bibnamefont{Kidder}},
  \bibnamefont{et~al.}, \bibinfo{journal}{Phys.Rev.}
  \textbf{\bibinfo{volume}{D84}}, \bibinfo{pages}{124052}
  (\bibinfo{year}{2011}), \eprint{1106.1021}.

\bibitem[{\citenamefont{Ajith et~al.}(2008)}]{Ajith:2007kx}
\bibinfo{author}{\bibfnamefont{P.}~\bibnamefont{Ajith}} \bibnamefont{et~al.},
  \bibinfo{journal}{Phys. Rev.} \textbf{\bibinfo{volume}{D77}},
  \bibinfo{pages}{104017} (\bibinfo{year}{2008}), \eprint{0710.2335}.

\bibitem[{\citenamefont{Ajith}(2008)}]{Ajith:2007xh}
\bibinfo{author}{\bibfnamefont{P.}~\bibnamefont{Ajith}},
  \bibinfo{journal}{Class. Quant. Grav.} \textbf{\bibinfo{volume}{25}},
  \bibinfo{pages}{114033} (\bibinfo{year}{2008}).

\bibitem[{\citenamefont{Ajith et~al.}(2011)\citenamefont{Ajith, Hannam, Husa,
  Chen, Br\"ugmann, Dorband, M\"uller, Ohme, Pollney, Reisswig
  et~al.}}]{Ajith:2009bn}
\bibinfo{author}{\bibfnamefont{P.}~\bibnamefont{Ajith}},
  \bibinfo{author}{\bibfnamefont{M.}~\bibnamefont{Hannam}},
  \bibinfo{author}{\bibfnamefont{S.}~\bibnamefont{Husa}},
  \bibinfo{author}{\bibfnamefont{Y.}~\bibnamefont{Chen}},
  \bibinfo{author}{\bibfnamefont{B.}~\bibnamefont{Br\"ugmann}},
  \bibinfo{author}{\bibfnamefont{N.}~\bibnamefont{Dorband}},
  \bibinfo{author}{\bibfnamefont{D.}~\bibnamefont{M\"uller}},
  \bibinfo{author}{\bibfnamefont{F.}~\bibnamefont{Ohme}},
  \bibinfo{author}{\bibfnamefont{D.}~\bibnamefont{Pollney}},
  \bibinfo{author}{\bibfnamefont{C.}~\bibnamefont{Reisswig}},
  \bibnamefont{et~al.}, \bibinfo{journal}{Phys. Rev. Lett.}
  \textbf{\bibinfo{volume}{106}}, \bibinfo{pages}{241101}
  (\bibinfo{year}{2011}), \eprint{0909.2867}.

\bibitem[{\citenamefont{Sintes and Vecchio}(1999)}]{Sintes:1999cg}
\bibinfo{author}{\bibfnamefont{A.~M.} \bibnamefont{Sintes}} \bibnamefont{and}
  \bibinfo{author}{\bibfnamefont{A.}~\bibnamefont{Vecchio}}
  (\bibinfo{year}{1999}), \eprint{gr-qc/0005058}.

\bibitem[{\citenamefont{Van Den~Broeck and
  Sengupta}(2007{\natexlab{a}})}]{VanDenBroeck:2006ar}
\bibinfo{author}{\bibfnamefont{C.}~\bibnamefont{Van Den~Broeck}}
  \bibnamefont{and} \bibinfo{author}{\bibfnamefont{A.~S.}
  \bibnamefont{Sengupta}}, \bibinfo{journal}{Class. Quant. Grav.}
  \textbf{\bibinfo{volume}{24}}, \bibinfo{pages}{1089}
  (\bibinfo{year}{2007}{\natexlab{a}}), \eprint{gr-qc/0610126}.

\bibitem[{\citenamefont{Van Den~Broeck and
  Sengupta}(2007{\natexlab{b}})}]{VanDenBroeck:2006qu}
\bibinfo{author}{\bibfnamefont{C.}~\bibnamefont{Van Den~Broeck}}
  \bibnamefont{and} \bibinfo{author}{\bibfnamefont{A.~S.}
  \bibnamefont{Sengupta}}, \bibinfo{journal}{Class.Quant.Grav.}
  \textbf{\bibinfo{volume}{24}}, \bibinfo{pages}{155}
  (\bibinfo{year}{2007}{\natexlab{b}}), \eprint{gr-qc/0607092}.

\bibitem[{\citenamefont{Cho et~al.}(2013)\citenamefont{Cho, Ochsner,
  O'Shaughnessy, Kim, and Lee}}]{Cho:2012ed}
\bibinfo{author}{\bibfnamefont{H.-S.} \bibnamefont{Cho}},
  \bibinfo{author}{\bibfnamefont{E.}~\bibnamefont{Ochsner}},
  \bibinfo{author}{\bibfnamefont{R.}~\bibnamefont{O'Shaughnessy}},
  \bibinfo{author}{\bibfnamefont{C.}~\bibnamefont{Kim}}, \bibnamefont{and}
  \bibinfo{author}{\bibfnamefont{C.-H.} \bibnamefont{Lee}},
  \bibinfo{journal}{Phys.Rev.} \textbf{\bibinfo{volume}{D87}},
  \bibinfo{pages}{024004} (\bibinfo{year}{2013}), \eprint{1209.4494}.

\bibitem[{\citenamefont{O'Shaughnessy
  et~al.}(2014{\natexlab{a}})\citenamefont{O'Shaughnessy, Farr, Ochsner, Cho,
  Kim et~al.}}]{O'Shaughnessy:2013vma}
\bibinfo{author}{\bibfnamefont{R.}~\bibnamefont{O'Shaughnessy}},
  \bibinfo{author}{\bibfnamefont{B.}~\bibnamefont{Farr}},
  \bibinfo{author}{\bibfnamefont{E.}~\bibnamefont{Ochsner}},
  \bibinfo{author}{\bibfnamefont{H.-S.} \bibnamefont{Cho}},
  \bibinfo{author}{\bibfnamefont{C.}~\bibnamefont{Kim}}, \bibnamefont{et~al.},
  \bibinfo{journal}{Phys.Rev.} \textbf{\bibinfo{volume}{D89}},
  \bibinfo{pages}{064048} (\bibinfo{year}{2014}{\natexlab{a}}),
  \eprint{1308.4704}.

\bibitem[{\citenamefont{O'Shaughnessy
  et~al.}(2014{\natexlab{b}})\citenamefont{O'Shaughnessy, Farr, Ochsner, Cho,
  Raymond et~al.}}]{O'Shaughnessy:2014dka}
\bibinfo{author}{\bibfnamefont{R.}~\bibnamefont{O'Shaughnessy}},
  \bibinfo{author}{\bibfnamefont{B.}~\bibnamefont{Farr}},
  \bibinfo{author}{\bibfnamefont{E.}~\bibnamefont{Ochsner}},
  \bibinfo{author}{\bibfnamefont{H.}~\bibnamefont{Cho}},
  \bibinfo{author}{\bibfnamefont{V.}~\bibnamefont{Raymond}},
  \bibnamefont{et~al.} (\bibinfo{year}{2014}{\natexlab{b}}),
  \eprint{1403.0544}.

\bibitem[{\citenamefont{Arun et~al.}(2009)\citenamefont{Arun, Mishra, Broeck,
  Iyer, Sathyaprakash et~al.}}]{Arun:2008xf}
\bibinfo{author}{\bibfnamefont{K.}~\bibnamefont{Arun}},
  \bibinfo{author}{\bibfnamefont{C.}~\bibnamefont{Mishra}},
  \bibinfo{author}{\bibfnamefont{C.~V.~D.} \bibnamefont{Broeck}},
  \bibinfo{author}{\bibfnamefont{B.}~\bibnamefont{Iyer}},
  \bibinfo{author}{\bibfnamefont{B.}~\bibnamefont{Sathyaprakash}},
  \bibnamefont{et~al.}, \bibinfo{journal}{Class.Quant.Grav.}
  \textbf{\bibinfo{volume}{26}}, \bibinfo{pages}{094021}
  (\bibinfo{year}{2009}), \eprint{0810.5727}.

\bibitem[{\citenamefont{Arun et~al.}(2007)\citenamefont{Arun, Iyer,
  Sathyaprakash, Sinha, and Broeck}}]{Arun:2007hu}
\bibinfo{author}{\bibfnamefont{K.}~\bibnamefont{Arun}},
  \bibinfo{author}{\bibfnamefont{B.~R.} \bibnamefont{Iyer}},
  \bibinfo{author}{\bibfnamefont{B.}~\bibnamefont{Sathyaprakash}},
  \bibinfo{author}{\bibfnamefont{S.}~\bibnamefont{Sinha}}, \bibnamefont{and}
  \bibinfo{author}{\bibfnamefont{C.~V.~D.} \bibnamefont{Broeck}},
  \bibinfo{journal}{Phys.Rev.} \textbf{\bibinfo{volume}{D76}},
  \bibinfo{pages}{104016} (\bibinfo{year}{2007}), \eprint{0707.3920}.

\bibitem[{\citenamefont{Pekowsky et~al.}(2013)\citenamefont{Pekowsky, Healy,
  Shoemaker, and Laguna}}]{Pekowsky:2013hm}
\bibinfo{author}{\bibfnamefont{L.}~\bibnamefont{Pekowsky}},
  \bibinfo{author}{\bibfnamefont{J.}~\bibnamefont{Healy}},
  \bibinfo{author}{\bibfnamefont{D.}~\bibnamefont{Shoemaker}},
  \bibnamefont{and} \bibinfo{author}{\bibfnamefont{P.}~\bibnamefont{Laguna}},
  \bibinfo{journal}{Phys. Rev. D} \textbf{\bibinfo{volume}{87}},
  \bibinfo{pages}{084008} (\bibinfo{year}{2013}).

\bibitem[{\citenamefont{Brown et~al.}(2013)\citenamefont{Brown, Kumar, and
  Nitz}}]{Brown:2013hm}
\bibinfo{author}{\bibfnamefont{D.~A.} \bibnamefont{Brown}},
  \bibinfo{author}{\bibfnamefont{P.}~\bibnamefont{Kumar}}, \bibnamefont{and}
  \bibinfo{author}{\bibfnamefont{A.~H.} \bibnamefont{Nitz}},
  \bibinfo{journal}{Phys. Rev. D} \textbf{\bibinfo{volume}{87}},
  \bibinfo{pages}{082004} (\bibinfo{year}{2013}).

\bibitem[{\citenamefont{Capano et~al.}(2013)\citenamefont{Capano, Pan, and
  Buonanno}}]{Capano:2013hm}
\bibinfo{author}{\bibfnamefont{C.}~\bibnamefont{Capano}},
  \bibinfo{author}{\bibfnamefont{Y.}~\bibnamefont{Pan}}, \bibnamefont{and}
  \bibinfo{author}{\bibfnamefont{A.}~\bibnamefont{Buonanno}}
  (\bibinfo{year}{2013}), \eprint{1311.1286}.

\bibitem[{\citenamefont{Littenberg et~al.}(2013)\citenamefont{Littenberg,
  Baker, Buonanno, and Kelly}}]{Littenberg:2012uj}
\bibinfo{author}{\bibfnamefont{T.~B.} \bibnamefont{Littenberg}},
  \bibinfo{author}{\bibfnamefont{J.~G.} \bibnamefont{Baker}},
  \bibinfo{author}{\bibfnamefont{A.}~\bibnamefont{Buonanno}}, \bibnamefont{and}
  \bibinfo{author}{\bibfnamefont{B.~J.} \bibnamefont{Kelly}},
  \bibinfo{journal}{Phys.Rev.} \textbf{\bibinfo{volume}{D87}},
  \bibinfo{pages}{104003} (\bibinfo{year}{2013}), \eprint{1210.0893}.

\bibitem[{\citenamefont{Ossokine et~al.}(2013)\citenamefont{Ossokine, Kidder,
  and Pfeiffer}}]{Ossokine:2013zga}
\bibinfo{author}{\bibfnamefont{S.}~\bibnamefont{Ossokine}},
  \bibinfo{author}{\bibfnamefont{L.~E.} \bibnamefont{Kidder}},
  \bibnamefont{and} \bibinfo{author}{\bibfnamefont{H.~P.}
  \bibnamefont{Pfeiffer}}, \bibinfo{journal}{Phys.Rev.}
  \textbf{\bibinfo{volume}{D88}}, \bibinfo{pages}{084031}
  (\bibinfo{year}{2013}), \eprint{1304.3067}.

\bibitem[{\citenamefont{Hemberger et~al.}(2013)\citenamefont{Hemberger, Scheel,
  Kidder, Szilagyi, Lovelace et~al.}}]{Hemberger:2012jz}
\bibinfo{author}{\bibfnamefont{D.~A.} \bibnamefont{Hemberger}},
  \bibinfo{author}{\bibfnamefont{M.~A.} \bibnamefont{Scheel}},
  \bibinfo{author}{\bibfnamefont{L.~E.} \bibnamefont{Kidder}},
  \bibinfo{author}{\bibfnamefont{B.}~\bibnamefont{Szilagyi}},
  \bibinfo{author}{\bibfnamefont{G.}~\bibnamefont{Lovelace}},
  \bibnamefont{et~al.}, \bibinfo{journal}{Class.Quant.Grav.}
  \textbf{\bibinfo{volume}{30}}, \bibinfo{pages}{115001}
  (\bibinfo{year}{2013}), \eprint{1211.6079}.

\bibitem[{\citenamefont{Szilagyi et~al.}(2009)\citenamefont{Szilagyi, Lindblom,
  and Scheel}}]{Szilagyi:2009qz}
\bibinfo{author}{\bibfnamefont{B.}~\bibnamefont{Szilagyi}},
  \bibinfo{author}{\bibfnamefont{L.}~\bibnamefont{Lindblom}}, \bibnamefont{and}
  \bibinfo{author}{\bibfnamefont{M.~A.} \bibnamefont{Scheel}},
  \bibinfo{journal}{Phys.Rev.} \textbf{\bibinfo{volume}{D80}},
  \bibinfo{pages}{124010} (\bibinfo{year}{2009}), \eprint{0909.3557}.

\bibitem[{\citenamefont{Boyle and Mroue}(2009)}]{Boyle:2009vi}
\bibinfo{author}{\bibfnamefont{M.}~\bibnamefont{Boyle}} \bibnamefont{and}
  \bibinfo{author}{\bibfnamefont{A.~H.} \bibnamefont{Mroue}},
  \bibinfo{journal}{Phys.Rev.} \textbf{\bibinfo{volume}{D80}},
  \bibinfo{pages}{124045} (\bibinfo{year}{2009}), \eprint{0905.3177}.

\bibitem[{\citenamefont{Scheel et~al.}(2009)\citenamefont{Scheel, Boyle, Chu,
  Kidder, Matthews et~al.}}]{Scheel:2008rj}
\bibinfo{author}{\bibfnamefont{M.~A.} \bibnamefont{Scheel}},
  \bibinfo{author}{\bibfnamefont{M.}~\bibnamefont{Boyle}},
  \bibinfo{author}{\bibfnamefont{T.}~\bibnamefont{Chu}},
  \bibinfo{author}{\bibfnamefont{L.~E.} \bibnamefont{Kidder}},
  \bibinfo{author}{\bibfnamefont{K.~D.} \bibnamefont{Matthews}},
  \bibnamefont{et~al.}, \bibinfo{journal}{Phys.Rev.}
  \textbf{\bibinfo{volume}{D79}}, \bibinfo{pages}{024003}
  (\bibinfo{year}{2009}), \eprint{0810.1767}.

\bibitem[{\citenamefont{Boyle et~al.}(2007)}]{Boyle:2007ft}
\bibinfo{author}{\bibfnamefont{M.}~\bibnamefont{Boyle}} \bibnamefont{et~al.},
  \bibinfo{journal}{Phys. Rev.} \textbf{\bibinfo{volume}{D76}},
  \bibinfo{pages}{124038} (\bibinfo{year}{2007}), \eprint{0710.0158}.

\bibitem[{\citenamefont{Scheel et~al.}(2006)\citenamefont{Scheel, Pfeiffer,
  Lindblom, Kidder, Rinne et~al.}}]{Scheel:2006gg}
\bibinfo{author}{\bibfnamefont{M.~A.} \bibnamefont{Scheel}},
  \bibinfo{author}{\bibfnamefont{H.~P.} \bibnamefont{Pfeiffer}},
  \bibinfo{author}{\bibfnamefont{L.}~\bibnamefont{Lindblom}},
  \bibinfo{author}{\bibfnamefont{L.~E.} \bibnamefont{Kidder}},
  \bibinfo{author}{\bibfnamefont{O.}~\bibnamefont{Rinne}},
  \bibnamefont{et~al.}, \bibinfo{journal}{Phys.Rev.}
  \textbf{\bibinfo{volume}{D74}}, \bibinfo{pages}{104006}
  (\bibinfo{year}{2006}), \eprint{gr-qc/0607056}.

\bibitem[{\citenamefont{Lindblom et~al.}(2006)\citenamefont{Lindblom, Scheel,
  Kidder, Owen, and Rinne}}]{Lindblom:2005qh}
\bibinfo{author}{\bibfnamefont{L.}~\bibnamefont{Lindblom}},
  \bibinfo{author}{\bibfnamefont{M.~A.} \bibnamefont{Scheel}},
  \bibinfo{author}{\bibfnamefont{L.~E.} \bibnamefont{Kidder}},
  \bibinfo{author}{\bibfnamefont{R.}~\bibnamefont{Owen}}, \bibnamefont{and}
  \bibinfo{author}{\bibfnamefont{O.}~\bibnamefont{Rinne}},
  \bibinfo{journal}{Class.Quant.Grav.} \textbf{\bibinfo{volume}{23}},
  \bibinfo{pages}{S447} (\bibinfo{year}{2006}), \eprint{gr-qc/0512093}.

\bibitem[{\citenamefont{Pfeiffer et~al.}(2003)\citenamefont{Pfeiffer, Kidder,
  Scheel, and Teukolsky}}]{Pfeiffer:2002wt}
\bibinfo{author}{\bibfnamefont{H.~P.} \bibnamefont{Pfeiffer}},
  \bibinfo{author}{\bibfnamefont{L.~E.} \bibnamefont{Kidder}},
  \bibinfo{author}{\bibfnamefont{M.~A.} \bibnamefont{Scheel}},
  \bibnamefont{and} \bibinfo{author}{\bibfnamefont{S.~A.}
  \bibnamefont{Teukolsky}}, \bibinfo{journal}{Comput.Phys.Commun.}
  \textbf{\bibinfo{volume}{152}}, \bibinfo{pages}{253} (\bibinfo{year}{2003}),
  \eprint{gr-qc/0202096}.

\bibitem[{SpE()}]{SpEC}
\bibinfo{note}{The Spectral Einstein Code},
  \urlprefix\url{http://www.black-holes.org/SpEC.html}.

\bibitem[{\citenamefont{Mroue and Pfeiffer}(2012)}]{Mroue:2012kv}
\bibinfo{author}{\bibfnamefont{A.~H.} \bibnamefont{Mroue}} \bibnamefont{and}
  \bibinfo{author}{\bibfnamefont{H.~P.} \bibnamefont{Pfeiffer}}
  (\bibinfo{year}{2012}), \eprint{1210.2958}.

\bibitem[{\citenamefont{Mroue et~al.}(2013)\citenamefont{Mroue, Scheel,
  Szilagyi, Pfeiffer, Boyle et~al.}}]{Mroue:2013xna}
\bibinfo{author}{\bibfnamefont{A.~H.} \bibnamefont{Mroue}},
  \bibinfo{author}{\bibfnamefont{M.~A.} \bibnamefont{Scheel}},
  \bibinfo{author}{\bibfnamefont{B.}~\bibnamefont{Szilagyi}},
  \bibinfo{author}{\bibfnamefont{H.~P.} \bibnamefont{Pfeiffer}},
  \bibinfo{author}{\bibfnamefont{M.}~\bibnamefont{Boyle}}, \bibnamefont{et~al.}
  (\bibinfo{year}{2013}), \eprint{1304.6077}.

\bibitem[{\citenamefont{Buchman et~al.}(2012)\citenamefont{Buchman, Pfeiffer,
  Scheel, and Szilagyi}}]{Buchman:2012dw}
\bibinfo{author}{\bibfnamefont{L.~T.} \bibnamefont{Buchman}},
  \bibinfo{author}{\bibfnamefont{H.~P.} \bibnamefont{Pfeiffer}},
  \bibinfo{author}{\bibfnamefont{M.~A.} \bibnamefont{Scheel}},
  \bibnamefont{and} \bibinfo{author}{\bibfnamefont{B.}~\bibnamefont{Szilagyi}},
  \bibinfo{journal}{Phys.Rev.} \textbf{\bibinfo{volume}{D86}},
  \bibinfo{pages}{084033} (\bibinfo{year}{2012}), \eprint{1206.3015}.

\bibitem[{SXS()}]{SXS-Catalog}
\bibinfo{note}{SXS Gravitational Waveform Database},
  \urlprefix\url{http://www.black-holes.org/waveforms/}.

\bibitem[{\citenamefont{Bruegmann et~al.}(2008)\citenamefont{Bruegmann,
  Gonzalez, Hannam, Husa, Sperhake et~al.}}]{Bruegmann:2006at}
\bibinfo{author}{\bibfnamefont{B.}~\bibnamefont{Bruegmann}},
  \bibinfo{author}{\bibfnamefont{J.~A.} \bibnamefont{Gonzalez}},
  \bibinfo{author}{\bibfnamefont{M.}~\bibnamefont{Hannam}},
  \bibinfo{author}{\bibfnamefont{S.}~\bibnamefont{Husa}},
  \bibinfo{author}{\bibfnamefont{U.}~\bibnamefont{Sperhake}},
  \bibnamefont{et~al.}, \bibinfo{journal}{Phys.Rev.}
  \textbf{\bibinfo{volume}{D77}}, \bibinfo{pages}{024027}
  (\bibinfo{year}{2008}), \eprint{gr-qc/0610128}.

\bibitem[{\citenamefont{Husa et~al.}(2008{\natexlab{a}})\citenamefont{Husa,
  Gonzalez, Hannam, Bruegmann, and Sperhake}}]{Husa:2007hp}
\bibinfo{author}{\bibfnamefont{S.}~\bibnamefont{Husa}},
  \bibinfo{author}{\bibfnamefont{J.~A.} \bibnamefont{Gonzalez}},
  \bibinfo{author}{\bibfnamefont{M.}~\bibnamefont{Hannam}},
  \bibinfo{author}{\bibfnamefont{B.}~\bibnamefont{Bruegmann}},
  \bibnamefont{and} \bibinfo{author}{\bibfnamefont{U.}~\bibnamefont{Sperhake}},
  \bibinfo{journal}{Class.Quant.Grav.} \textbf{\bibinfo{volume}{25}},
  \bibinfo{pages}{105006} (\bibinfo{year}{2008}{\natexlab{a}}),
  \eprint{0706.0740}.

\bibitem[{\citenamefont{Apostolatos}(1995)}]{Apostolatos:1995pj}
\bibinfo{author}{\bibfnamefont{T.~A.} \bibnamefont{Apostolatos}},
  \bibinfo{journal}{Phys. Rev. D} \textbf{\bibinfo{volume}{52}},
  \bibinfo{pages}{605} (\bibinfo{year}{1995}).

\bibitem[{\citenamefont{Vallisneri}(2008)}]{Vallisneri:2007ev}
\bibinfo{author}{\bibfnamefont{M.}~\bibnamefont{Vallisneri}},
  \bibinfo{journal}{Phys.Rev.} \textbf{\bibinfo{volume}{D77}},
  \bibinfo{pages}{042001} (\bibinfo{year}{2008}), \eprint{gr-qc/0703086}.

\bibitem[{\citenamefont{Lovelace et~al.}(2008)\citenamefont{Lovelace, Owen,
  Pfeiffer, and Chu}}]{Lovelace:2008tw}
\bibinfo{author}{\bibfnamefont{G.}~\bibnamefont{Lovelace}},
  \bibinfo{author}{\bibfnamefont{R.}~\bibnamefont{Owen}},
  \bibinfo{author}{\bibfnamefont{H.~P.} \bibnamefont{Pfeiffer}},
  \bibnamefont{and} \bibinfo{author}{\bibfnamefont{T.}~\bibnamefont{Chu}},
  \bibinfo{journal}{Phys.Rev.} \textbf{\bibinfo{volume}{D78}},
  \bibinfo{pages}{084017} (\bibinfo{year}{2008}), \eprint{0805.4192}.

\bibitem[{\citenamefont{Pfeiffer et~al.}(2007)\citenamefont{Pfeiffer, Brown,
  Kidder, Lindblom, Lovelace et~al.}}]{Pfeiffer:2007yz}
\bibinfo{author}{\bibfnamefont{H.~P.} \bibnamefont{Pfeiffer}},
  \bibinfo{author}{\bibfnamefont{D.~A.} \bibnamefont{Brown}},
  \bibinfo{author}{\bibfnamefont{L.~E.} \bibnamefont{Kidder}},
  \bibinfo{author}{\bibfnamefont{L.}~\bibnamefont{Lindblom}},
  \bibinfo{author}{\bibfnamefont{G.}~\bibnamefont{Lovelace}},
  \bibnamefont{et~al.}, \bibinfo{journal}{Class.Quant.Grav.}
  \textbf{\bibinfo{volume}{24}}, \bibinfo{pages}{S59} (\bibinfo{year}{2007}),
  \eprint{gr-qc/0702106}.

\bibitem[{\citenamefont{Caudill et~al.}(2006)\citenamefont{Caudill, Cook,
  Grigsby, and Pfeiffer}}]{Caudill:2006hw}
\bibinfo{author}{\bibfnamefont{M.}~\bibnamefont{Caudill}},
  \bibinfo{author}{\bibfnamefont{G.~B.} \bibnamefont{Cook}},
  \bibinfo{author}{\bibfnamefont{J.~D.} \bibnamefont{Grigsby}},
  \bibnamefont{and} \bibinfo{author}{\bibfnamefont{H.~P.}
  \bibnamefont{Pfeiffer}}, \bibinfo{journal}{Phys.Rev.}
  \textbf{\bibinfo{volume}{D74}}, \bibinfo{pages}{064011}
  (\bibinfo{year}{2006}), \eprint{gr-qc/0605053}.

\bibitem[{\citenamefont{Cook and Pfeiffer}(2004)}]{Cook:2004kt}
\bibinfo{author}{\bibfnamefont{G.~B.} \bibnamefont{Cook}} \bibnamefont{and}
  \bibinfo{author}{\bibfnamefont{H.~P.} \bibnamefont{Pfeiffer}},
  \bibinfo{journal}{Phys.Rev.} \textbf{\bibinfo{volume}{D70}},
  \bibinfo{pages}{104016} (\bibinfo{year}{2004}), \eprint{gr-qc/0407078}.

\bibitem[{\citenamefont{Friedrich}(1996)}]{Friedrich:1996hq}
\bibinfo{author}{\bibfnamefont{H.}~\bibnamefont{Friedrich}},
  \bibinfo{journal}{Class.Quant.Grav.} \textbf{\bibinfo{volume}{13}},
  \bibinfo{pages}{1451} (\bibinfo{year}{1996}).

\bibitem[{\citenamefont{Garfinkle}(2002)}]{Garfinkle:2001ni}
\bibinfo{author}{\bibfnamefont{D.}~\bibnamefont{Garfinkle}},
  \bibinfo{journal}{Phys.Rev.} \textbf{\bibinfo{volume}{D65}},
  \bibinfo{pages}{044029} (\bibinfo{year}{2002}), \eprint{gr-qc/0110013}.

\bibitem[{\citenamefont{Pretorius}(2005)}]{Pretorius:2004jg}
\bibinfo{author}{\bibfnamefont{F.}~\bibnamefont{Pretorius}},
  \bibinfo{journal}{Class.Quant.Grav.} \textbf{\bibinfo{volume}{22}},
  \bibinfo{pages}{425} (\bibinfo{year}{2005}), \eprint{gr-qc/0407110}.

\bibitem[{\citenamefont{Brandt and Bruegmann}(1997)}]{Brandt:1997tf}
\bibinfo{author}{\bibfnamefont{S.}~\bibnamefont{Brandt}} \bibnamefont{and}
  \bibinfo{author}{\bibfnamefont{B.}~\bibnamefont{Bruegmann}},
  \bibinfo{journal}{Phys.Rev.Lett.} \textbf{\bibinfo{volume}{78}},
  \bibinfo{pages}{3606} (\bibinfo{year}{1997}), \eprint{gr-qc/9703066}.

\bibitem[{\citenamefont{Bowen and York}(1980)}]{Bowen:1980yu}
\bibinfo{author}{\bibfnamefont{J.~M.} \bibnamefont{Bowen}} \bibnamefont{and}
  \bibinfo{author}{\bibfnamefont{J.}~\bibnamefont{York},
  \bibfnamefont{James~W.}}, \bibinfo{journal}{Phys.Rev.}
  \textbf{\bibinfo{volume}{D21}}, \bibinfo{pages}{2047} (\bibinfo{year}{1980}).

\bibitem[{\citenamefont{Ansorg et~al.}(2004)\citenamefont{Ansorg, Bruegmann,
  and Tichy}}]{Ansorg:2004ds}
\bibinfo{author}{\bibfnamefont{M.}~\bibnamefont{Ansorg}},
  \bibinfo{author}{\bibfnamefont{B.}~\bibnamefont{Bruegmann}},
  \bibnamefont{and} \bibinfo{author}{\bibfnamefont{W.}~\bibnamefont{Tichy}},
  \bibinfo{journal}{Phys.Rev.} \textbf{\bibinfo{volume}{D70}},
  \bibinfo{pages}{064011} (\bibinfo{year}{2004}), \eprint{gr-qc/0404056}.

\bibitem[{\citenamefont{Husa et~al.}(2008{\natexlab{b}})\citenamefont{Husa,
  Hannam, Gonzalez, Sperhake, and Bruegmann}}]{Husa:2007rh}
\bibinfo{author}{\bibfnamefont{S.}~\bibnamefont{Husa}},
  \bibinfo{author}{\bibfnamefont{M.}~\bibnamefont{Hannam}},
  \bibinfo{author}{\bibfnamefont{J.~A.} \bibnamefont{Gonzalez}},
  \bibinfo{author}{\bibfnamefont{U.}~\bibnamefont{Sperhake}}, \bibnamefont{and}
  \bibinfo{author}{\bibfnamefont{B.}~\bibnamefont{Bruegmann}},
  \bibinfo{journal}{Phys.Rev.} \textbf{\bibinfo{volume}{D77}},
  \bibinfo{pages}{044037} (\bibinfo{year}{2008}{\natexlab{b}}),
  \eprint{0706.0904}.

\bibitem[{\citenamefont{Hannam et~al.}(2010)\citenamefont{Hannam, Husa, Ohme,
  Muller, and Bruegmann}}]{Hannam:2010ec}
\bibinfo{author}{\bibfnamefont{M.}~\bibnamefont{Hannam}},
  \bibinfo{author}{\bibfnamefont{S.}~\bibnamefont{Husa}},
  \bibinfo{author}{\bibfnamefont{F.}~\bibnamefont{Ohme}},
  \bibinfo{author}{\bibfnamefont{D.}~\bibnamefont{Muller}}, \bibnamefont{and}
  \bibinfo{author}{\bibfnamefont{B.}~\bibnamefont{Bruegmann}},
  \bibinfo{journal}{Phys. Rev.} \textbf{\bibinfo{volume}{D82}},
  \bibinfo{pages}{124008} (\bibinfo{year}{2010}), \eprint{1007.4789}.

\bibitem[{\citenamefont{P{\"u}rrer et~al.}(2012)\citenamefont{P{\"u}rrer, Husa,
  and Hannam}}]{Purrer:2012wy}
\bibinfo{author}{\bibfnamefont{M.}~\bibnamefont{P{\"u}rrer}},
  \bibinfo{author}{\bibfnamefont{S.}~\bibnamefont{Husa}}, \bibnamefont{and}
  \bibinfo{author}{\bibfnamefont{M.}~\bibnamefont{Hannam}},
  \bibinfo{journal}{Phys.Rev.} \textbf{\bibinfo{volume}{D85}},
  \bibinfo{pages}{124051} (\bibinfo{year}{2012}), \eprint{1203.4258}.

\bibitem[{\citenamefont{Campanelli et~al.}(2006)\citenamefont{Campanelli,
  Lousto, Marronetti, and Zlochower}}]{Campanelli:2005dd}
\bibinfo{author}{\bibfnamefont{M.}~\bibnamefont{Campanelli}},
  \bibinfo{author}{\bibfnamefont{C.~O.} \bibnamefont{Lousto}},
  \bibinfo{author}{\bibfnamefont{P.}~\bibnamefont{Marronetti}},
  \bibnamefont{and}
  \bibinfo{author}{\bibfnamefont{Y.}~\bibnamefont{Zlochower}},
  \bibinfo{journal}{Phys. Rev. Lett.} \textbf{\bibinfo{volume}{96}},
  \bibinfo{pages}{111101} (\bibinfo{year}{2006}), \eprint{gr-qc/0511048}.

\bibitem[{\citenamefont{Baker et~al.}(2006)\citenamefont{Baker, Centrella,
  Choi, Koppitz, and van Meter}}]{Baker:2005vv}
\bibinfo{author}{\bibfnamefont{J.~G.} \bibnamefont{Baker}},
  \bibinfo{author}{\bibfnamefont{J.}~\bibnamefont{Centrella}},
  \bibinfo{author}{\bibfnamefont{D.-I.} \bibnamefont{Choi}},
  \bibinfo{author}{\bibfnamefont{M.}~\bibnamefont{Koppitz}}, \bibnamefont{and}
  \bibinfo{author}{\bibfnamefont{J.}~\bibnamefont{van Meter}},
  \bibinfo{journal}{Phys. Rev. Lett.} \textbf{\bibinfo{volume}{96}},
  \bibinfo{pages}{111102} (\bibinfo{year}{2006}), \eprint{gr-qc/0511103}.

\bibitem[{\citenamefont{Hannam et~al.}(2007)\citenamefont{Hannam, Husa,
  Pollney, Br{\"u}gmann, and {\'O Murchadha}}}]{Hannam:2006vv}
\bibinfo{author}{\bibfnamefont{M.}~\bibnamefont{Hannam}},
  \bibinfo{author}{\bibfnamefont{S.}~\bibnamefont{Husa}},
  \bibinfo{author}{\bibfnamefont{D.}~\bibnamefont{Pollney}},
  \bibinfo{author}{\bibfnamefont{B.}~\bibnamefont{Br{\"u}gmann}},
  \bibnamefont{and} \bibinfo{author}{\bibfnamefont{N.}~\bibnamefont{{\'O
  Murchadha}}}, \bibinfo{journal}{Phys. Rev. Lett.}
  \textbf{\bibinfo{volume}{99}}, \bibinfo{pages}{241102}
  (\bibinfo{year}{2007}), \eprint{gr-qc/0606099}.

\bibitem[{\citenamefont{Shibata and Nakamura}(1995)}]{Shibata:1995we}
\bibinfo{author}{\bibfnamefont{M.}~\bibnamefont{Shibata}} \bibnamefont{and}
  \bibinfo{author}{\bibfnamefont{T.}~\bibnamefont{Nakamura}},
  \bibinfo{journal}{Phys. Rev.} \textbf{\bibinfo{volume}{D52}},
  \bibinfo{pages}{5428} (\bibinfo{year}{1995}).

\bibitem[{\citenamefont{Baumgarte and Shapiro}(1999)}]{Baumgarte:1998te}
\bibinfo{author}{\bibfnamefont{T.~W.} \bibnamefont{Baumgarte}}
  \bibnamefont{and} \bibinfo{author}{\bibfnamefont{S.~L.}
  \bibnamefont{Shapiro}}, \bibinfo{journal}{Phys. Rev.}
  \textbf{\bibinfo{volume}{D59}}, \bibinfo{pages}{024007}
  (\bibinfo{year}{1999}), \eprint{gr-qc/9810065}.

\bibitem[{\citenamefont{Reisswig and Pollney}(2011)}]{Reisswig:2010di}
\bibinfo{author}{\bibfnamefont{C.}~\bibnamefont{Reisswig}} \bibnamefont{and}
  \bibinfo{author}{\bibfnamefont{D.}~\bibnamefont{Pollney}},
  \bibinfo{journal}{Class.Quant.Grav.} \textbf{\bibinfo{volume}{28}},
  \bibinfo{pages}{195015} (\bibinfo{year}{2011}), \eprint{1006.1632}.

\bibitem[{\citenamefont{Ajith et~al.}(2012)\citenamefont{Ajith, Boyle, Brown,
  Brügmann, Buchman et~al.}}]{Ajith:2012az}
\bibinfo{author}{\bibfnamefont{P.}~\bibnamefont{Ajith}},
  \bibinfo{author}{\bibfnamefont{M.}~\bibnamefont{Boyle}},
  \bibinfo{author}{\bibfnamefont{D.~A.} \bibnamefont{Brown}},
  \bibinfo{author}{\bibfnamefont{B.}~\bibnamefont{Brügmann}},
  \bibinfo{author}{\bibfnamefont{L.~T.} \bibnamefont{Buchman}},
  \bibnamefont{et~al.}, \bibinfo{journal}{Class.Quant.Grav.}
  \textbf{\bibinfo{volume}{29}}, \bibinfo{pages}{124001}
  (\bibinfo{year}{2012}), \eprint{1201.5319}.

\bibitem[{\citenamefont{Hinder et~al.}(2014)\citenamefont{Hinder, Buonanno,
  Boyle, Etienne, Healy et~al.}}]{Hinder:2013oqa}
\bibinfo{author}{\bibfnamefont{I.}~\bibnamefont{Hinder}},
  \bibinfo{author}{\bibfnamefont{A.}~\bibnamefont{Buonanno}},
  \bibinfo{author}{\bibfnamefont{M.}~\bibnamefont{Boyle}},
  \bibinfo{author}{\bibfnamefont{Z.~B.} \bibnamefont{Etienne}},
  \bibinfo{author}{\bibfnamefont{J.}~\bibnamefont{Healy}},
  \bibnamefont{et~al.}, \bibinfo{journal}{Class.Quant.Grav.}
  \textbf{\bibinfo{volume}{31}}, \bibinfo{pages}{025012}
  (\bibinfo{year}{2014}), \eprint{1307.5307}.

\bibitem[{\citenamefont{Blanchet et~al.}(2008)\citenamefont{Blanchet, Faye,
  Iyer, and Sinha}}]{Blanchet:2008je}
\bibinfo{author}{\bibfnamefont{L.}~\bibnamefont{Blanchet}},
  \bibinfo{author}{\bibfnamefont{G.}~\bibnamefont{Faye}},
  \bibinfo{author}{\bibfnamefont{B.~R.} \bibnamefont{Iyer}}, \bibnamefont{and}
  \bibinfo{author}{\bibfnamefont{S.}~\bibnamefont{Sinha}},
  \bibinfo{journal}{Class.Quant.Grav.} \textbf{\bibinfo{volume}{25}},
  \bibinfo{pages}{165003} (\bibinfo{year}{2008}), \eprint{0802.1249}.

\bibitem[{\citenamefont{Blanchet et~al.}(2004)\citenamefont{Blanchet, Damour,
  Esposito-Farese, and Iyer}}]{Blanchet:2004ek}
\bibinfo{author}{\bibfnamefont{L.}~\bibnamefont{Blanchet}},
  \bibinfo{author}{\bibfnamefont{T.}~\bibnamefont{Damour}},
  \bibinfo{author}{\bibfnamefont{G.}~\bibnamefont{Esposito-Farese}},
  \bibnamefont{and} \bibinfo{author}{\bibfnamefont{B.~R.} \bibnamefont{Iyer}},
  \bibinfo{journal}{Phys. Rev. Lett.} \textbf{\bibinfo{volume}{93}},
  \bibinfo{pages}{091101} (\bibinfo{year}{2004}), \eprint{gr-qc/0406012}.

\bibitem[{\citenamefont{Ajith et~al.}(2007)}]{Ajith:2007qp}
\bibinfo{author}{\bibfnamefont{P.}~\bibnamefont{Ajith}} \bibnamefont{et~al.},
  \bibinfo{journal}{Class. Quant. Grav.} \textbf{\bibinfo{volume}{24}},
  \bibinfo{pages}{S689} (\bibinfo{year}{2007}).

\bibitem[{lal()}]{lalsimulation}
\bibinfo{note}{LALSimulation is part of the LALSuite software package},
  \urlprefix\url{https://www.lsc-group.phys.uwm.edu/daswg/projects/lalsuite.html}.

\bibitem[{adl()}]{adligo-psd}
\emph{\bibinfo{title}{Advanced ligo anticipated sensitivity curves}},
  \bibinfo{note}{{LIGO} Document T0900288-v3},
  \urlprefix\url{https://dcc.ligo.org/LIGO-T0900288/public}.

\bibitem[{sci()}]{scipy}
\bibinfo{note}{The SciPy software library}, \urlprefix\url{http://scipy.org/}.

\bibitem[{\citenamefont{{LIGO Scientific Collaboration and Virgo
  Collaboration}}()}]{lvc-whitepaper-2014-2015}
\bibinfo{author}{\bibnamefont{{LIGO Scientific Collaboration and Virgo
  Collaboration}}}, \emph{\bibinfo{title}{The {LSC}-{V}irgo white paper on
  gravitational wave searches and astrophysics}}, \bibinfo{note}{{LIGO}
  Document Number LIGO-T1400054-v6}.

\bibitem[{\citenamefont{Cutler and Flanagan}(1994)}]{Cutler:1994ys}
\bibinfo{author}{\bibfnamefont{C.}~\bibnamefont{Cutler}} \bibnamefont{and}
  \bibinfo{author}{\bibfnamefont{E.~E.} \bibnamefont{Flanagan}},
  \bibinfo{journal}{Phys.Rev.} \textbf{\bibinfo{volume}{D49}},
  \bibinfo{pages}{2658} (\bibinfo{year}{1994}).

\bibitem[{\citenamefont{Poisson and Will}(1995)}]{Poisson:1995ef}
\bibinfo{author}{\bibfnamefont{E.}~\bibnamefont{Poisson}} \bibnamefont{and}
  \bibinfo{author}{\bibfnamefont{C.~M.} \bibnamefont{Will}},
  \bibinfo{journal}{Phys.Rev.} \textbf{\bibinfo{volume}{D52}},
  \bibinfo{pages}{848} (\bibinfo{year}{1995}), \eprint{gr-qc/9502040}.

\bibitem[{\citenamefont{Baird et~al.}(2013)\citenamefont{Baird, Fairhurst,
  Hannam, and Murphy}}]{Baird:2012cu}
\bibinfo{author}{\bibfnamefont{E.}~\bibnamefont{Baird}},
  \bibinfo{author}{\bibfnamefont{S.}~\bibnamefont{Fairhurst}},
  \bibinfo{author}{\bibfnamefont{M.}~\bibnamefont{Hannam}}, \bibnamefont{and}
  \bibinfo{author}{\bibfnamefont{P.}~\bibnamefont{Murphy}},
  \bibinfo{journal}{Phys.Rev.} \textbf{\bibinfo{volume}{D87}},
  \bibinfo{pages}{024035} (\bibinfo{year}{2013}), \eprint{1211.0546}.

\bibitem[{\citenamefont{Hannam et~al.}(2013)\citenamefont{Hannam, Brown,
  Fairhurst, Fryer, and Harry}}]{Hannam:2013uu}
\bibinfo{author}{\bibfnamefont{M.}~\bibnamefont{Hannam}},
  \bibinfo{author}{\bibfnamefont{D.~A.} \bibnamefont{Brown}},
  \bibinfo{author}{\bibfnamefont{S.}~\bibnamefont{Fairhurst}},
  \bibinfo{author}{\bibfnamefont{C.~L.} \bibnamefont{Fryer}}, \bibnamefont{and}
  \bibinfo{author}{\bibfnamefont{I.~W.} \bibnamefont{Harry}},
  \bibinfo{journal}{Astrophys.J.} \textbf{\bibinfo{volume}{766}},
  \bibinfo{pages}{L14} (\bibinfo{year}{2013}), \eprint{1301.5616}.

\bibitem[{\citenamefont{Chatziioannou et~al.}(2014)\citenamefont{Chatziioannou,
  Cornish, Klein, and Yunes}}]{Chatziioannou:2014coa}
\bibinfo{author}{\bibfnamefont{K.}~\bibnamefont{Chatziioannou}},
  \bibinfo{author}{\bibfnamefont{N.}~\bibnamefont{Cornish}},
  \bibinfo{author}{\bibfnamefont{A.}~\bibnamefont{Klein}}, \bibnamefont{and}
  \bibinfo{author}{\bibfnamefont{N.}~\bibnamefont{Yunes}}
  (\bibinfo{year}{2014}), \eprint{1402.3581}.

\end{thebibliography}

\end{document}